\documentclass[11pt]{article}

\usepackage{amssymb}
\usepackage{graphicx}
\usepackage{latexsym}
\usepackage{longtable}
\usepackage{lscape}
\usepackage{rotating}
\usepackage[sort&compress]{natbib}

\usepackage{epsfig,epic,amsfonts}
\usepackage{wrapfig}

\def\thefootnote{\fnsymbol{footnote}}

\newcommand{\eq}{\begin{equation}}
\newcommand{\en}{\end{equation}}
\newcommand{\be}{\begin{equation}}
\newcommand{\ee}{\end{equation}}
\newcommand{\eqa}{\begin{eqnarray}}
\newcommand{\ena}{\end{eqnarray}}
\newcommand{\ba}{\begin{eqnarray}}
\newcommand{\ea}{\end{eqnarray}}

\bibpunct{[}{]}{,}{n}{}{}

\begin{document}

\begin{titlepage}
\vskip0.5cm
\begin{flushright}
\hbox{}
\end{flushright}

\vskip1.0cm

\begin{center}
{\LARGE \bf Genome-Wide Survey of MicroRNA -}
\vskip0.3cm
{\LARGE \bf Transcription Factor Feed-Forward}
\vskip0.3cm
{\LARGE \bf Regulatory Circuits in Human}
\end{center}

\vskip1.3cm
\centerline{ Angela Re $^{a,\#}$, Davide Cor\'a $^{b,d\#}$, Daniela Taverna $^{c,d}$ and Michele~Caselle $^{b,d*}$}
\vskip1.0cm

\centerline{\sl a) CIBIO - Centre for Integrative Biology, University of Trento}
\centerline{\sl Via delle Regole 101, I-38100 Trento, Italy}
\centerline{\sl e--mail: \hskip 0.5cm re@science.unitn.it }

\vskip0.8cm

\centerline{\sl b) Department of Theoretical Physics, University of Torino and INFN}
\centerline{\sl Via Pietro~Giuria 1, I-10125 Torino, Italy}
\centerline{\sl e--mail: \hskip 0.5cm cora@to.infn.it, caselle@to.infn.it}

\vskip0.8cm

\centerline{\sl c) Department of Oncological Sciences, University of Torino and }
\centerline{\sl Molecular Biotechnology Center, Via Nizza 52, I-10126 Torino, Italy}
\centerline{\sl e--mail: \hskip 0.5cm daniela.taverna@unito.it}

\vskip0.8cm

\centerline{\sl d) Center for Complex Systems in Molecular Biology and Medicine, University of Torino }
\centerline{\sl Via Accademia Albertina 13, I-10100 Torino, Italy}

\vskip2.0cm
 
\# equal contribution\\

* corresponding author, email: caselle@to.infn.it, phone: +39-011-6707205, fax: +39-011-6707214

\end{titlepage}

\clearpage

{\bf Summary}

\vskip0.5cm

In this work we describe a computational framework for the genome-wide identification and characterization
of mixed transcriptional / post-transcriptional regulatory circuits in human. We concentrated in particular on
Feed-Forward Loops (FFL) in which a master transcription factor regulates a microRNA 
and together with it a set of joint target protein coding genes. The circuits were assembled  
with a two step procedure. 
We first constructed separately the transcriptional and post-transcriptional components
 of the human regulatory network
by looking for conserved overrepresented motifs in human and mouse promoters and 3'-UTRs. 
Then we combined the two subnetworks looking for mixed feed-forward regulatory interactions, finding a total of 638 putative (merged) FFLs. 
In order to investigate their biological relevance we filtered these circuits using
 three selection criteria: (I) GeneOntology enrichment among the joint
targets of the FFL (II) Independent computational evidence for the regulatory interactions of the FFL, 
extracted from external databases (III) Relevance of the FFL in cancer.
Most of the selected FFLs seem to be involved in  various aspects of organism 
development and differentiation. We finally discuss a few of the most interesting cases in detail. 

\vskip1.0cm


\vskip1.0cm

{\it Running title:} Human Regulatory Circuits.

\clearpage

\setcounter{footnote}{0}
\def\thefootnote{\arabic{footnote}}

\section*{Background}
\label{background}

     A basic notion of modern system biology is
     that biological functions are performed by groups of genes which act in
an interdipendent and synergic way.
     This is true in particular for regulatory processes for which it is by
now mandatory to assume
     a ``network'' point of view.

     Among the various important consequences of this approach a prominent
role is played by the notion
     of ``network motif''. The idea is that a complex network (say a
regulatory network) can be divided into
     simpler, distinct regulatory patterns called network motifs, typically
composed by three or four
     interacting components which are able to perform elementary signal
processing functions. Network motifs can
     be thougth of as the smallest functional modules of the network and, by
suitably combining them, the whole
     complexity of the original network can be recovered.\\

     In this paper we shall be interested in ``mixed'' network motifs
involving both transcriptional
     (T) and post-transcriptional (PT) regulatory interactions and in
particular we shall especially focus our attention on the mixed feed-forward
     loops. Feed-forward loops (FFLs) have been shown to be
     one of the most important  classes of transcriptional network
motifs~\citep{alon2002,alon2002b}.
     The major goal of our work is to extend them also including
post-transcriptional regulatory
     interactions.\\

        Indeed, in the last few years it has become more and more evident
that post-transcriptional processes play a
     much more important role than previously expected in the regulation of
gene expression.

Among the various mechanisms of post-transcriptional regulation a prominent
role is played by a class of small RNAs called
microRNAs (miRNAs), reviewed in \citep{he:2004,filipowicz}. miRNAs are a
family of $\sim$22nt small non-coding RNAs which
negatively regulate gene expression at the post-transcriptional level, in a
wide range of organisms. They are involved in different
biological functions, including, developmental timing, pattern formation,
embryogenesis, differentiation, organogenesis, growth control
and cell death. They certainly play a major role in human deseases as well
\citep{alvarez-garcia:2005, calin:2006}.\\

Mature miRNAs are produced from longer precursors which in some cases
cluster together in the so called miRNA ``Transcriptional Units''
(TU)~\citep{landgraf} and their expression is regulated by the same
molecular mechanisms which control protein-coding gene expression.
Even though the precise mechanism of action of the miRNAs is not well
understood, the current paradigm is that in animals
miRNAs are able to repress the translation of target genes by binding, in
general, in a Watson-Crick complementary manner
to 7 nucleotides (nts) long sequences present at the 3'-UnTranslated Region
(3'-UTR) of the regulated genes. The binding
usually involves nts 2 to 8 of the miRNA, the so called ``seed''. Often the
miRNA binding sites at the 3'-UTR of the
target genes are overrepresented \citep{lai:2002, lewis:2003, Nielsen2007,
krek:2005, lewis:2005, xie:2005, chan:2005}.  \\

All these findings, in addition with a large amount of work related to the
discovery of transcription factor binding sites
(for a recent review see for instance \citep{elnitski:2006}), suggest that
both transcriptional and post-transcriptional
regulatory interactions could be predicted in silico by searching
overrepresented short sequences of nts present
in promoters or  3'-UTRs and by filtering the results with suitable
evolutionary or functional constraints.

Stemming from these observations, the aim of our work was 
to use computational tools to generate a list of
feed-forward loops in which a master transcription
factor (TF) regulated a miRNA and together with it a set of  target genes
(see Figure 1a). We performed a genome wide ``ab initio'' search, and we found in this way 
a total of 638 putative (merged) FFLs. 
In order investigate their biological relevance we then filtered these circuits using
 three selection criteria: (I) GeneOntology enrichemnt among the joint
targets of the FFL. (II) Independent computational evidence for the regolatory interactions of the FFL, 
extracted from the ECRbase, miRBase, PicTar end TargetScan databases. (III) Relevance to cancer of the 
FFL as deduced from their intersection with the Oncomir and Cancer gene census databases.

In a few cases some (or all) of the regulatory interactions which composed
the feed-forward loop were found to be already known in
the literature, with their interplay in a closed regulatory circuit not
noticed, thus representing an important validation of our approach. 
However for several loops we predicted new regulatory interactions which represent 
reliable targets for experimental validation.

Let us finally notice that in this work we only discuss
 the simplest non trivial regulatory circuits (feed-forward
loops). However our raw data could be easily used to construct more complex network motifs. 
For this reason we let them accessible to the interested investigators
as Supplementary files.

\section*{Results}
\label{results}

Here we provide a collection of circuits that explicitly link a Transcription Factor (TF) and a microRNA (miRNA) which both regulate a
set of common target genes (Figure 1a). To this end we (1) constructed a transcriptional regulatory network; (2) defined a miRNA 
mediated post-transcriptional regulatory network; (3) merged the two networks and (4) filtered the results with
various selection criteria (Figure 2). In the next section we shall then discuss a few cases in more detail.

\subsection*{Circuits identification}

\subsubsection*{Construction of a human transcriptional regulatory network}

The starting point of our analysis was the construction of a database of promoter regions for both protein-coding 
and miRNA genes for human and mouse. Details of this construction are reported in the Materials and Methods section.
Here we only stress our main choices. For protein-coding genes we selected the core promoter region nearby the 
Transcription Start Site (TSS), whereas for the miRNA promoters we chose to merge together all the miRNAs present in the so called  
``transcriptional units'' (TUs) proposed in \citep{landgraf}, kept only the conserved TUs (human and mouse) 
and then selected the putative core promoter regions (see Supplementary files S1 and S2). \\

We then identified, separately for human and mouse, sets of genes (protein-coding plus miRNAs) sharing overrepresented
oligonucleotides (oligos), 6 to 9 nts long, in their associated promoter regions. Next we selected the oligos for which
the human and mouse sets contained a statistically significant fraction of orthologous genes. In doing so we used a 
binomial model for the assessment of overrepresentation and an alignment-free evolutionary methodology for the 
identification of conserved oligos, as previously used in \citep{cora:2005,cora:2007}. This approach was also extended 
to the putative promoters of miRNA genes. All the sequences were Repeat-Masked, and we took into account either redundancy 
due to superposition of the same genomic areas or protein-coding exons or correction for CG content of the sequences themselves.
As a final result, we ended up with a catalogue of cis-regulatory motifs conserved in the core promoter regions of human and mouse 
protein-coding or miRNA genes, each endowed with a score (the Pvalue of the evolutionary conservation test, described in
the Materials and Methods). We then applied corrections for multiple testing and ranking, setting 0.1 as 
False Discovery rate (FDR).\\ 

The last step was the association of the surving motifs with known transcription factor binding sequences (TFBSs), where 
possible, to obtain a list of putative TF--target gene interactions. To this end we used the TRANSFAC \citep{transfac:2006} 
database and the list of consensus motifs reported in~\citep{xie:2005}. 

Fixing 0.1 as FDR level, we obtained a catalogue of 2031 oligos that could be associated to known TFBSs 
for a total of 115 different TFs. These 2031 oligos targeted a total of 21159 genes (20972 
protein-coding and 187 miRNAs) and almost every gene in the Ensembl~\citep{birney:2007} database was present at least once 
in our network. In parallel to that our motif discovery procedure further identified 20216 significant motifs but for which
we were not able to make any strong association with known TFBSs consensus.

The dataset of associations between motifs and genes represents our transcriptional regulatory network and was 
the starting point for the circuits identification (see Supplementary files S3 and S4). A relevant role in the
following will be played by the subnetwork describing the transcriptional regulation of miRNAs. This subnetwork
involves 110 TFs (out of 115 of the whole network) targeting a total of 187 miRNAs (see Supplementary file S4).

\subsubsection*{Construction of a human post-transcriptional regulatory network} 

We used a very similar approach for the construction of the post-transcriptional regulatory 
network and used a dataset of 3'-UTRs for all the protein-coding genes in the human and mouse genomes.
We ended up with a catalogue of 3989 short oligos (in this case 7-mers) overrepresented and conserved in human
and mouse after corrections for multiple testing and ranking, again setting 0.1 as FDR threshold in our motifs discovery 
pipeline. Although the ab-initio unbiased procedure that we used could discover different kind of 
post-transcriptional regulatory 
motif \citep{cora:2007}, we kept only those motifs which could be associated with ``seeds'' of our known mature miRNAs (193 in total). 
182 out of 3989 motifs turned out to match with at least one seed present in 140 out of 193 mature miRNAs 
(in some cases the motif could be associated to  more than one miRNA). These motifs targeted a total of 17266 
protein-coding genes which represented our post-transcriptional regulatory networks reported in Supplementary file S5.\\

\subsubsection*{Construction of the human mixed feed-forward loops catalogue}	

Once equipped with these two regulatory networks, we could in principle integrate their complementary 
information in various different ways. Here we concentrated on the class of mixed Feed-Forward Loops (FFLs) discussed 
for instance  in \citep{hornstein:2006,tsang,shalgi} because biologically important and relatively simple
to relate to experimental evidences and validations. We integrated the two networks looking for all possible
cases in which a master TF regulates a miRNA and together with it a set of protein-coding Joint Targets (JT). Notice that, as
mentioned above, for each TF we associated all the motifs compatible with its binding site and its variants
as they are reported in
the TRANSFAC \citep{transfac:2006} and in the ~\citep{xie:2005} collections. 
In this way the intrinsic variability of regulatory binding sites, apparently neglected by our 
method,
since we used fixed motifs, was restored in the final results.

We were able to obtain a list of 5030 different ``single target circuits'', each of them defined by a 
single TF as master regulator,
a single mature miRNA and a single protein-coding joint target. 
We then grouped together all the single target circuits
sharing the same pair of TF and  miRNA and obtained as final result,
638 ``merged'' circuits, each composed by a known TF acting as master regulator, a mature 
miRNA and a list of protein-coding joint targets (see Figure 1a). 
These 638 circuits involved a total of 2625 joint target genes, 101 transcription factors and 
133 miRNAs. The number of joint targets in these circuits ranged from 1 to 38. 74\% of the 
circuits 
targeted up to 10 genes.

The raw data relative to these circuits can be found in Supplementary file S6.\\  

Besides the motifs used to build the above described circuits we have several other
cis-regulatory
upstream motifs in our transcriptional networks which could not be related to a known TFBS. 
These motifs can be considered as new, putative, regulatory
sequences \citep{cora:2005} and, even if we are not able to associate a precise TF (or any other kind of regulatory mechanism)
to them we decided to extend the above construction also to these sequences. 
In these cases it would be too difficult to reconstruct the variability of the binding site for the 
corresponding putative unknown TF, so we decided to construct only the FFL 
in which the exact same unidentified 
and fixed motif was present in the upstream region of both the target protein-coding gene and the
co-regulating miRNA and, as above, closed the loop only if the target gene
was also a target of the considered miRNA.  

In this way we obtained 4035 different circuits which 
included various motifs with different sizes on the promoter regions: 
170, 6 nts long; 128, 7 nts long; 440, 8 nts long; 3297, 9 nts long. 
The number of joint targets in these circuits,
after merging on the same cis-regulatory motifs, ranged from 1 to 5. 79\% of the circuits targeted 
one single gene.

All the raw data concerning these fixed-motif circuits can be found in Supplementary file S7.\\

\subsubsection*{Circuits assessment I: functional analysis.}


As a first way to select biologically relevant FFL among our results
we analyzed each one of the 638 merged circuits looking for an enrichment
in Gene Ontology categories in the set of their joint targets. To assess this enrichment
we used the standard exact Fisher test with a p-value threshold  $p<10^{-4}$.
Previous experience on similar enrichment tests~\citep{cora:2004,cora:2005}
shows that this is a rather robust way to keep into account multiple testing of GO 
categories which, being highly correlated, cannot be treated with a standard Bonferroni correction.
Details regarding this analysis are available in Materials and Methods. 

As a final result of this analysis, we end with a list of 32 merged mixed feed-forward loops
 (corresponding to 380 
single-target FFLs). These circuits 
involve a total of 344 joint target
protein-coding genes, 24 TFs and 25 mature miRNAs. We report in Table 1 a selected list of such 
loops with a subset of the most representative Gene Ontology enriched annotations; the complete list of results
is available in Supplementary file S8. \\

\subsubsection*{Circuits assessment II: comparison with existing computational databases.}

To further assess the relevance of the circuits that we identified 
 we developed an annotation scheme, based on the existence of additional computational 
evidences for each circuit link. 
To this end we used  ECRbase~\citep{ovcharenko:2007} and the data collected in~\citep{bino_john} 
for the transcriptional links and
the miRBase \citep{john:2004}, PicTar \citep{krek:2005} and TargetScan \citep{lewis:2003} databases for the
post-transcriptional ones. Let us see in more detail how we used these sources of information:
\begin{itemize}
\item
The Evolutionary Conserved Regions database (ECRbase~\citep{ovcharenko:2007})
is a collection of evolutionary conserved regions, promoters and TFBSs in 
Vertebrate genomes, based on genome-wide alignments created mainly with the Blastz program.  Even if both our
pipeline and ECRbase are based on evolutionary conservation, this ingredient is implemented in a very different 
way in the two approaches. ECRbase looks for conserved blocks identified via whole-genome alignments, while 
we implemented evolutionary conservation using an alignment-free approach. In this way we were able to validate 
216 TF--target gene links and 98 TF--miRNA links.\\

\item
Ref.\citep{bino_john} is a computational study of miRNA biogenesis.
The regulatory interactions reported in \citep{bino_john} 
are  of particular interest for our assessment procedure
since their pipeline is very different from ours. With this tool we were able to validate 
64 TF--miRNA links. It is interesting to notice that these 64 miRNAs were controlled by
only 9 transcription factors (the important role of these ``hub'' TFs was already
 noticed in \citep{bino_john})

\item
The miRBase \citep{john:2004}, PicTar \citep{krek:2005} and TargetScan \citep{lewis:2003} databases are by now
an accepted standard in the miRNA literature. They are based on strategies which are definitely different from
our 
pipeline and are somehow complementary in their approaches.
In this way we were able to validate the miRNA--target gene link for 607 circuits (503 by miRBase, 343 by PicTar and 560 by
TargetScan).\\
 \end{itemize}
 
The results of these comparisons are summarized in Table 2a, while Table 3 reports the top ten TFs 
ranked by out-degree and the top-ten miRNAs scored by in-degree.\\

In Table 2b we report a selection of a few circuits which turned out to be validated by the above tests. 
In the Supplementary file S8 one can find the complete list of results.

\subsubsection*{Circuits assessment III: looking for cancer related FFLs.}

In these last few years it is becoming increasingly clear that miRNAs play a central role in cancer 
development. About half of the human miRNAs are located in cancer-related chromosomal regions and 
miRNA expression profiling correlates with various cancers and it is used to improve cancer diagnosis. 
This supports the definition of a subset of miRNAs as ``oncomiRs'' \citep{OncomiRs:Kerscher}.\\

We filtered our results looking for circuits containing at least one cancer-related miRNA or target gene. 
To identify cancer related genes we used the list of oncomiRs reported in \citep{OncomiRs:Kerscher} and 
\citep{OncomiRs:Zhang} while for the protein-coding target genes we compiled a list of genes showing 
mutations in cancer based on the Cancer Gene Census catalogue.\\

In particular we found 24 circuits in which at least {\it two} cancer-related genes 
(e.g. an oncomiR and a target or a TF and an oncomiR) were present (see Table 4). 
The full list of cancer-related circuits  is available in the Supplementary
file S9 and S10.

\section*{Discussion}
\label{discussion}



\subsection*{Potential function of mixed feed-forward circuits}

Depending on the type of transcriptional regulation (excitatory or inhibitory) exerted by the master TF
on the miRNA and on the targets, the Feed-Forward Loops that we study in this paper may be classified  (following \citep{hornstein:2006})
as  coherent, if the master TF and the miRNA act in a coherent way
on the target, or incoherent in the opposite case. A similar classification can be found in \citep{tsang} where the two classes of FFL
were named as Type II or Type I respectively (see Figure 3, in which we chose to follow the same notations of \citep{tsang})
Due to the computational procedure that we adopted to identify the FFLs, based on sequence
analysis only, we were not able to recognize if the action of the master TF was excitatory or inhibitory and thus if the FFL that we 
obtained was of Type I or Type II. Accordingly in Figures 1 and 4 we avoided to identify the links which connect the 
master TF to its targets as excitatory or inhibitory and used a different notation. 
Obviously the two types of circuits may lead to very different 
behaviours~\citep{hornstein:2006,tsang}. Type II (coherent) circuits lead to a reinforcement of the transcriptional regulation at 
the post-transcriptional level and might be important to eliminate the already transcribed mRNAs when the transcription
of a target gene is switched off. Type I can be used to stabilize the steady state production of a protein by dumping 
transcriptional fluctuations. In a simple TF-target interaction any fluctuation of master TF
could induce a non-linear increase in the amount of its target products. The presence, among the targets, of a miRNA 
which downregulates the other targets might represent a simple and effective way to control these fluctuations. 
Another interesting possibility (discussed for instance in~\citep{shalgi}) occurs if a temporal gap exists between 
the activation of the target gene and the miRNA repressor. This could be the case, for instance, if the binding sequences 
of the master TF in the two promoters have different affinities or if there is a delay in the miRNA maturation process.
In this case the Type I circuit could be used to express the target protein within a well defined time window. 
In this respect it is interesting to observe that one of the most studied mixed FFL is a Type I-like: here the role of 
master TF is played by c-Myc, which induces the expression of miR-17-5p and miR-20a and also of the joint target, 
E2F1, which, in turn, is repressed by the same miRNAs~\citep{mendell:2005}. Needless to say that these elementary 
FFLs when embedded in more complex circuits can lead to more sophisticated behaviours 
(see for instance the discussion in~\citep{tsang,shalgi}, and in particular ~\citep{vergassola:2005}). 
We shall see below an example of this type of construction.

\subsection*{Analysis of the mixed feed-forward circuits in terms of network motifs}


Elementary regulatory circuits (the so called "network motifs")
were shown to be over-represented in transcriptional networks~\citep{alon2002, alon2002b}. This very interesting
observation led a few
authors to conjecture that functionally important network motifs should always be over-represented and to use
this criterion as a tool to identify them.  
This assumption is somehow controversial and  is currently challenged 
by some other authors e.g. ~\citep{vergassola:2005,konagurthu:2008}. Our data represent a perfect setting
to test this over-representation conjecture.

In order to quantify the over-representation we perfomed a set of randomization tests. 
Results are reported in Supplementary Figure S1 and details are available in the Supporting Text. 
Briefly, we carried out three type of randomizations:

\begin{itemize}

\item {\bf Random reshuffling of miRNA promoters and seeds} We rebuilt the entire database of mixed feed-forward circuits (i.e. the entire
      pipeline designed in Figure 2), but using randomly shuffled versions of the miRNA promoters and random sets of
      7-mers as miRNA seeds. The principle of this procedure was to perform the same analysis of correlation between
      transcriptional and post-transcriptional regulatory networks but considering as connection between the two regulatory layers a randomized
      version of the real known miRNAs, in terms of their in-degree (the miRNA promoter) and out-degree (the miRNA seed). 

\item {\bf Edge switching} We applied a randomization strategy on the real transcriptional and post-transcriptional
      regulatory network obtained with our pipeline, similar to the one used in \citep{martinez:2008}. The Edge switching 
      strategy is able to randomize the real network, preserving the individual degree of each node in the network.

\item {\bf Complete node replacement} We applied a second, more drastic, randomization strategy 
on the real transcriptional 
and post-transcriptional
      regulatory network obtained with our pipeline, in this case with no constraint on the randomization
      procedure \citep{martinez:2008}.
      
\end{itemize}

The results reported in the Supplementary Figure S1 (panel A) show that for the three randomization strategies, 
the number of circuits recognized in the real regulatory network is statistically higher than the one 
found in the random versions
(Random reshuffling of miRNA promoters and seeds: $Z = 3.5$, Edge switching: $Z = 8.3$, Complete node replacement: $Z = 8.9$).
However it is important to notice that the actual number of mixed feed-forward loops identified 
in the randomized versions of the regulatory network is always rather large. Thus, even if the
over-representation is
statistically significant it would be very inefficient (i.e. it would lead to a large number of false positive
identifications) to use it 
as the only tool to identify functionally relevant mixed FFL.
Interestingly, our results are in good agreement with a similar analysis reported in 
\citep{martinez:2008}. This is particularly significant since our approach and that of \citep{martinez:2008} for
the identification of TF and miRNA regulatory interactions  are
totally different.
In \citep{martinez:2008}, the authors presented the first genome-scale Caenorhabditis 
elegans miRNA regulatory network 
that contains experimentally mapped transcriptional TF $\rightarrow$ miRNA interactions, as well as computationally predicted 
post-transcriptional miRNA $\rightarrow$ TF connections. They then looked at the properties of mixed feedback loops, comparing
their findings with network randomizations: the average number of loops in randomized networks was 
always about half the number of real loops they identified.

\subsection*{Analysis of the Gene Ontology enrichment results.}

In the Supplementary file S8 we report a detailed view of the  GO 
enrichment results both at the level of joint target sets and 
of single gene analysis. 
Besides the intrinsic interest of several of these annotations it is interesting to observe
that the set of GO categories enriched in our circuits shows somehow a general trend.

We observe overrepresentation of GO terms describing several aspects 
of organism development such as {\it differentiation}, {\it proliferation}, {\it apoptosis}, 
{\it programmed cell death} and {\it cellular migration}. These results are in good agreement with
the predictions about the biological meanings of the feed-forward loops reported in  \citep{hornstein:2006}. 
Specifically, our data provide evidence for functions of several circuits 
in the cardiac and skeletal, neural and hematopoietic cell lineages.  

A similar pattern emerges if we look at the single-gene enrichment analysis. {\it Multi-cellular organisms development},
{\it cell differentiation}, {\it cell proliferation} and {\it apoptosis} directly annotate respectively 108, 56 and 48 
target genes included in the annotated circuits.

Finally it is interesting to notice that several circuits seem to be involved, according to the GO analysis, in basal 
mechanisms of post-translational regulation such as {\it protein amino acid phosphorylation} and
in the {\it ubiquitin cycle} (with as much as 57 annotated genes).

All these observations agree with the idea  that the mixed (T-PT) motifs and in particular the feed-forward loops
that we discuss in this paper play a fundamental role in all those processes (like tissue development and cell
differentiation) which are characterized by a high
degree of complexity and require the simultaneous fine tuning of several different players. Strikigly, it is worth noting
that this result was obtained here with a completely ab-initio bioinformatics sequence analysis strategy. 

\subsection*{Comparison of our results with the database \citep{Zeller2006} 
of ChIP-PET c-Myc targets.}

Besides the above tests, in order to evaluate the reliability of our transcriptional regulatory 
network we compared our results  with a set of c-Myc targets reported in 
\citep{Zeller2006}. This database contains a
  a genome-wide, unbiased characterization of direct Myc binding targets in a model 
  of human B lymphoid tumor using ChIP coupled with pair-end ditag sequencing 
  analysis (ChIP-PET) and reports a total of 2088 targets. 
  
  The choice of the c-Myc TF is not random. Besides being a very interesting TF
  it is present in several of our FFLs and as such it plays a central role in the transcriptional side of our regulatory network.
  In particular the first example that we shall discuss below contains c-Myc as master TF.
  
  Looking at the intersection between the 2088 targets of \citep{Zeller2006} and the 1979
  predicted by our analysis we found 253 targets in common corresponding to a p-value
  of $1.1\times 10^{-6}$ (Fisher test). This result is even more impressive if compared with the number of
  intersections of the Zeller dataset with the list of c-Myc targets reported in
  the TRANSFAC database \citep{transfac:2006}. Out of 235 TRANSFAC targets only 27 were present in Zeller's dataset corresponding
  to a p value of 0.21.

  As a further test we performed the same comparison for the transcriptional network 
  obtained choosing as promoter the (-500/+100) region around the TSS. In this case we found 1612 putative c-Myc targets 
  in which 203 were in common with the dataset of ref. \citep{Zeller2006} corresponding to a slightly higher p-value
  $p=8.4\times 10^{-5}$.

\subsection*{Dependence of our results on the choice of the promoter's region.}
  In the construction of the transcriptional regulatory network we chose to consider 
  the interval (-900/+100) around the TSS for the promoter regions. 
  In order to test the dependence on this choice of our results we performed the same analysis choosing 
  as promoter region the interval (-500/+100) around the TSS. This is somehow an extreme choice and represents a very
  stringent test of the robustness of our network. 
  Looking at the mixed FFL we found a total of 6682 "single target" FFLs (to be
   compared with the 5030 of the (-900/+100) case) in which  1769 were in common with the (-900/+100) run.
   Remarkably enough all the circuits which we discussed in the text (and more generally most
   of the circuits surviving our assessment tests) turned out to  be present in both releases. The
   complete list of circuits obtained in the (-500/+100) is reported in the Supplementary file S11.
     
   In order to complete this analysis we also performed
   the randomization tests and the comparison with the c-Myc database discussed above for the (-500/+100) FFLs. 
   We found comparable results with those obtained
   in the (-900/+100) case:  the number of circuits of  the real regulatory network turned out to be statistically higher than the ones 
   found in the random simulations. In particular for the the first two tests we found an improvement of the $Z$ values while for
   the third one we found a slightly worse values of $Z$. All these results are reported in the second panel of Supplementary Figure S1.
   Also for the c-Myc analysis we found results comparable with those obtained in the (-900/+100) case, with a slight worsening of the 
   p-value of the intersection.
   More precisely the c-Myc targets in the (-500/+100) transcriptional network turned out
   to be 1612 of which 203 in common with the Zeller c-Myc dataset corresponding to a p-value of $8\times 10^{-5}$.

   We consider all these findings 
   as an indication of the overall robustness of our results. 

\subsection*{Comparison with related works}
Mixed T-PT regulatory circuits have been recently studied in two interesting papers~\citep{tsang,shalgi}. 
It is worthwhile to compare their results with our analysis which is similar in spirit, but slightly more
complete in the final results.

In \citep{tsang} the authors studied various types of feed-forward and feedback loops involving miRNAs, 
their target genes and transcriptional regulators as a tool to explain
the (anti-)correlations between the expression levels of 
miRNAs and of their target genes. This study was grounded on a predicted miRNA-mediated network and 
did not include the transcriptional regulatory network of miRNAs that was unavailable at that time. 
Hence, to the best of our knowledge, no actual explicit loops was determined (see also \citep{martinez:2008}).

In \citep{shalgi} the authors used pre-compiled TF- and miRNA-mediated networks and studied global 
and local properties 
of the two networks separately. Additionally, they provided a catalogue of network designs in the 
co-regulated network, 
including feed-forward loops. Both the TF- and the miRNA-mediated networks 
in  \citep{shalgi} were 
obtained from sequence-based identification of regulatory features in promoters and 3'-UTRs. 
This makes the \citep{shalgi} 
study more comparable to ours than \citep{tsang}. For this reason we decided to perform a more detailed
comparison with our results.

Unfortunately, this study did not report explicitly the circuits
(including joint target genes) but only provided a list of 16 pairs of
co-regulating TFs and miRNAs involved in feed-forward loops.
We obtained these pairs using as input the PSSMs (Position Specific Scoring
Matrices) and microRNAs listed in the Supplementary Table S3 of \citep{shalgi} and
then mapping the
PSSMs to the corresponding TFs.

It turns out that none of these predictions are contained in our dataset. A detailed comparison of the two pipelines
shows that there are a few important reasons behind this disagreement:

\begin{itemize}

\item Different annotation for mature miRNA identifiers due to the older miRBase release used in  \citep{shalgi} (8.2 
	versus 9.2): e.g. pairs involve miR-10 in \citep{shalgi} while miRBase 9.2 reports miR-10a and miR-10b; similarly 
	for miR-142 and miR-142-5p,-3p.

\item Different assignment of mature miRNAs to pre-miRNAs: e.g. in  \citep{shalgi} the authors assign miR-7 to mir-7-1 
	while miRBase 9.2 assigns miR-7 to mir-7-3.

\item Different organization of pre-miRNAs in transcriptional units: in \citep{shalgi} miRNA are clustered in precursors 
according to physical proximity while we rely on human/mouse conserved transcriptional units reported in \citep{landgraf}.

\item Different definition of miRNA promoters: \citep{shalgi} uses 10kb upstream of the 5' most pre-miRNA for each cluster 
while we use 1kb upstream of the 5' most pre-miRNA for each transcriptional unit.

\item Different solutions for predicted transcription factor binding sites: \citep{shalgi} uses PSSMs from TRANSFAC 
release 8.3, and using pre-compiled lists of interactions available in the UCSC hg17 genome assembly, while we mainly 
map ab-initio conserved and overrepresented motifs to transcription factor binding sites. 

\item Different solutions for predicted mature miRNA binding sites: \citep{shalgi} uses TargetScan (release 3.0) and 
PicTar (picTarMiRNA4Way track in the UCSC genome browser) while we map conserved and overrepresented motifs in 3'-UTRs 
to mature miRNAs by means of miRBase release 9.2.

\end{itemize}

As a final comment on this comparison let us stress that probably
one of the major novelties of the present analysis with respect to existing works 
is the particular attention we payed
to the definition of miRNA promoters and in the search of their putative binding sequences.
Accordingly, besides the final list of FFLs, we consider as one of our most interesting results
the subset of our transcriptional regulatory network involving miRNAs as targets. 
This subnetwork includes a total of 110 TFs targeting 187 miRNAs and is reported in the Supplementary file S4.

\subsection*{Description of a few interesting circuits}

As a final part of this section let us discuss in more detail the biological relevance of a few of our results.
We have chosen to discuss a few examples for each of the three assessment pipelines.  

We first present a case in which our pipeline is able to predict circuits already known
in the literature and for which all the links are experimentally validated: this is the case
of the circuits involving c-Myc as master TF and hsa-miR-17-5p and hsa-miR-20a as
post-transcriptional regulators. In particular, one of the predicted joint target genes results
to be the E2F1 gene, closing in this way the circuit exactly on the target gene experimentally 
assessed and used as major example in the discussion of ~\citep{hornstein:2006}.

In the remaining examples some (or all) 
the genes embedded in the circuits were already annotated to related 
functions in the literature but their combination in a closed FFL was not noticed. 
We consider these cases as further succesful validations of our approach. 

\begin{itemize}

\item {\bf The c-Myc, hsa-miR-20a/miR-17-5p circuit}

In this circuit c-Myc is the master TF and hsa-miR-20a the post-transcriptional regulator. This circuit
contains 11 joint targets among which the, E2F1. The complete list of joint
targets is reported in Table 1. The FFL involving E2F1 is well known in the literature. It was discussed for the
first time in~\citep{mendell:2005} and  is expected to play a role in the control of cell proliferation, growth
and apoptosis.
With our analysis we could identify several other genes sharing the same regulatory pattern of E2F1 and we
expect that at least some of them could be involved in the same biological processes. In this respect it is
interesting to find among the other targets NFAT5 which is known to play
a critical role in heart, vasculature, muscle and
nervous tissue development. Similarly, it seems interesting to find  MAPK1  which, like E2F1, is an
anti-apoptotic gene. These observations could suggest a similar functional role also for the
remaining joint targets.

This circuit also allows us to discuss how our data could be used  to obtain more complex regulatory motifs.
Combining different entries of our databses it is easy to find a circuit 
involving, besides c-Myc, hsa-miR-20a and E2F1,
also E2F2 which turns out to be simultaneously targeted by E2F1 and by  hsa-miR-20a (see Figure 4). This is a
rather non trivial result since it 
is well known that different TFs of the E2F family tend to act together in a
concerted way. We see in this example a simple network motif in which this cooperative action is present and 
is tightly regulated.

\item {\bf The AREB6, hsa-miR-375 circuit} 

One of the most interesting entries of Table 1 is the feed-forward loop which involves
the transcriptional repressor zinc-finger E-box binding homeobox 1 AREB6 (also known as ZEB1), hsa-miR-375 and a set
of 14 joint target genes. Owing to the following observations, we surmise its function in
embryonic development and physiology of pancreas.
ZEB1 is a crucial inducer of the embryonic program 'epithelial-mesenchymal transition' (EMT)
that facilitates tissue remodeling during embryonic development.
miR-375 is essential for embryonic pancreatic islet development as well as for
endocrine pancreas function  where it was demonstrated to regulate the process of
exocytosis of insulin during glucose-stimulated insulin release \citep{Joglekar}.
Notably GO analysis globally annotates the set of target genes to patterning in
embryonic development which is consistent with the regulatory roles of ZEB-1 and miR-375.
Moreover the hypothesis of a function in insulin secretion is
strengthened by the following observations: \citep{gershengorn} reports strong
evidence that EMT can provide cells for replacement therapy in diabetes; among
the target genes,  HNF1$\beta$ (also known as TCF2) is responsible for MODY \citep{gudmundsson}, a
form of diabetes characterized by defective insulin secretion of pancreatic $\beta$-cells.

\item{ \bf The MEF-2, hsa-miR-133a circuit}

This is one of the entries of Table 1. It contains only 2 joint targets: BRUNOL4 and PLP2, but the presence of
BRUNOL4 turns out to be highly non trivial. In fact the  myocyte enhancing factor-2 (MEF-2), 
hsa-miR-133a and the RNA-binding protein BRUNOL4 
have been shown to control altogether cardiomyocyte hypertrophy. In this case, it is also possible to 
envisage a feed-back effect because cardiac repression of BRUONOL4 activity disrupts alternative 
splicing of MEF-2 and leads to cardiac hypertrophy \citep{Ladd:2003}.
Finally it is important to stress that the regulatory interaction between MEF-2 and  hsa-mir-133a  which we
predicted with our in silico analysis
was indeed observed experimentally in~\citep{liu}. 

\item{ \bf The C-REL , hsa-mir-199a circuit}

Another interesting circuit relates C-REL, a member of the NFKB family, and miR-199a. MiR-199a has been identified 
as a miRNA signature in human ovarian cancer. MiR-199a down-modulation in epithelial ovarian cells is 
reported in \cite{Iorio2007} and, interestingly, miR-199a has lately been shown to affect NFKB activity 
in ovarian cancer cells \cite{Chen2008}. Among the joint targets for this circuit, let us mention: DDR1, 
a receptor tyrosine kinase, whose expression is restricted to epithelial cells and significantly high in 
epithelial ovarian cells; Sp2 that is a transcriptional repressor of the tumor suppressor gene CEACAM1in 
epithelial cells \cite{Phan2004}.

\item{ \bf The HSF2 , hsa-let-7f circuit}

Looking at the cancer related list one of the most interestin entries is the one which
relates the transcription factor HSF2 and the hsa-let-7f miRNA. 
The DNA-binding protein heat shock factor-2 (HSF2) and hsa-let-7f jointly regulate 
a number of target genes such as MYCN, ESPL1, PLSCR3, PDCD4, MTO1 and FMO2. 
Several observations point to an involvement of this circuit in cell cycle 
progression with relevant implications in cancer.
HSF2 role in cancer is being elucidated \citep{Wilkerson} by the observation 
that it functions as bookmarking factor not only for heat shock responsive 
genes but also for genes that are involved in regulation of cell apoptosis 
and proliferation (such as Hsp90, Hsp27 and c-Fos).
Among target genes, the MYCN oncogene is crucial in neuronal development 
and its amplification is currently the only molecular marker adopted in neuroblastoma 
clinical treatments. The MYC family oncogenes are known to deregulate cell cycle 
progression, apoptosis and genomic instability. In neuroblastoma cell lines, 
N-Myc can induce genomic instability by centrosome amplification. 
Interestingly, HSF2 and hsa-let-7f regulate the extra spindle poles 
like-1 (ESPL1) that mediates mitotic sister chromatid segregation. 
The programmed cell death-4 (PDCD4) is also linked to progression through cell cycle by 
mediating MAPK kinase activity and JNK activity. The phospholipid scramblase-3 (PLSCR3) 
is a mitochondrial integrator of apoptotic signals. Interestingly also the mitochondrial 
translation optimization-1 homolog (MTO1) and the flavin containing monooxygenase-2 (FMO2) 
promote local effects on mitochondria. 
Finally, MYCN has recently been reported as a direct target of miR-34a. 
Here we add that let-7f targets MYCN. Notably let-7f belongs to the let-7 family of oncomiRs 
and, in particular, let-7f has been found involved in cell aging \citep{Wagner}.

\end{itemize}

As a final remark, we would like to stress that interesting convergence of cooperative biological 
functions can be observed also in circuits in which we were not able to identify a putative
master TF and therefore were not processed with our assesment pipeline. As an example let us 
mention the UST gene (Ensembl id: ENSG00000111962), which is involved in heparan sulfate-dependent
growth factor signaling during myogenesis and in ion buffering; UST linked to hsa-miR-1
(see Supplementary file S7) which in turn promotes skeletal muscle proliferation and 
differentiation and is involved in heart electrical conductions as well~\citep{srivastava}.

\section*{Conclusions}
\label{Conclusions}

The main purpose of this work was to sistematically investigate connections between
transcriptional and post-transcriptional network interactions, in the human genome. 
To this end, we designed a bioinformatic pipeline, mainly
based on sequence analysis of human and mouse genomes, that is able to
costruct, in particular, a catalogue of mixed Feed-Forward Loops (FFLs)
in which a master transcription factor regulates a miRNA and,
together with it, a set of joint target protein-coding genes. These circuits
were then prioritized based on various selection criteria. We also analyzed a few of them in detail looking for
a possible biological role. 
The lists of FFLs selected in this way are the
major results of our work and our findings demonstrate in particular a connection between such loops and aspects
of organisms development and differentiation. 
Moreover, one of the outcomes resulting from our study is the design 
of a putative TF regulatory network of human miRNA genes.

As a concluding remark it is important to stress that we consider the present work only as a first step along this research line.
For both technical and biological reasons it is likely that we miss several regulatory circuits in our network.
We  discussed in detail the technical issues and the related problems. Let us comment here on one of the main biological 
issues which should certainly be addressed in future works.  
One of our main assumptions is that we can associate a well defined promoter to a well defined gene. However several 
recent studies on the widespread presence of alternative splicing and
Transcription Start Sites (TSS) (see for instance \citep{Pan2008})
show that this is probably a too restrictive choice. Moreover alternatively spliced isoforms of the same gene may have completely different functions 
and play different roles in the regulatory network.  More generally the notion of ``gene'' by itself
is experiencing a deep redefinition in the last years (\citep{Pesole2008}).
Notwithstanding this, the good agreement that we found with some existing experimental data suggests that our approach may represent a reliable step 
toward a better understanding of gene regolatory networks
 and in particular it could give some useful insight on the complex interplay of their transcriptional
and post-transcriptional layers.

\section*{Materials and Methods}
\label{methods}

\subsection*{miRNA transcriptional units}

We obtained genomic coordinates of human and mouse pre-miRNA hairpins from the miRBase
\citep{Griffiths-Jones:2006} miRNA sequence database (release 9.2). Consistently, human and mouse 
protein-coding genes and annotations were obtained from the Ensembl database \citep{birney:2007}, 
release 46, corresponding 
to the human genome assembly hg18 and to the mouse genome assembly mmu8.
Mapping of pre-miRNAs to overlapping protein-coding genes was performed 
using the mirGen database
(\verb+http://www.diana.pcbi.upenn.edu/miRGen/v3/+)
which provided us with a list of all the pre-miRNA hairpins which overlapped 
to annotated genes and gave the precise location of the pre-miRNA 
hairpin within the gene. In this study, from the Ensembl database we selected only 
protein-coding genes labelled as ``KNOWN'', for both human and mouse. 
Pre-miRNAs were defined as genic if they were located within annotated exons, introns
 or flanking untranslated regions. miRNA hairpins were retained in our study only if they had an 
 orthologous copy in mouse. 
This selection was performed using the human-to-mouse 
orthology table compiled by \citep{landgraf} and provided as their Supplementary 
Table S15. 

An important role in our analysis is played by the notion of ``Transcriptional Units'' (TU) which are 
clusters of miRNA hairpins located in nearby positions along the DNA, and  supposed to be
transcribed together in a single poly-miRNA precursor~\citep{landgraf}.
Both cDNA and EST expression data~\citep{landgraf,enright} support the idea that
miRNAs belonging to the same TU are co-transcribed. For this reason we shall treat them as an unique (miRNA)
gene and associate the same promoter (the one corresponding to the Transcriptional Start Site, TSS, of the transcriptional unit)
to all the miRNAs belonging to the TU.

Taking together isolated miRNAs and TU we were able to identify a total of 130 miRNA 
precursors for the human genome and the corresponding 130 orthologues for the mouse genome. 
68 out of 130 were non-genic and 62 were located within a KNOWN gene. A direct 
inspection showed that 53 of these genic pre-miRNAs shared the same orientation with 
the host gene while the remaining 9 had the opposite orientation. These 130 precursors 
corresponded to a total of 193 mature miRNAs. These mature miRNAs and their ``seeds'' 
represented the list of input motifs for the target search algorithms and the bases 
of our discussions.

The list of TUs, their most 5'-upstream members, their genomic coordinates, their locations 
relative to protein-coding genes and additional orthology annotations can be found in 
Supplementary file S1, for human and  mouse. We then provide the corresponding mature miRNAs used in this
study in Supplementary file S2, for human and mouse.

\subsection*{Definition of promoter regions}

For the analysis of promoter regions, we prepared two distinct datasets, one for protein-coding genes 
and one for miRNA genes. All the sequences and annotations used  were extracted from the Ensembl database, 
version 46.\\

\emph{Protein-coding genes.}~~ We selected the complete list of protein-coding genes, for both human and
mouse, retaining only those labeled as ``KNOWN''. For each gene, we then selected only the longest transcript, 
again among those labeled as ``KNOWN''. For each of these genes, as putative promoter sequence we chose
the region starting from nt -900 upstream of the TSS and ending at nt +100 downstream of the TSS 
(being the TSS at position +1) of the selected transcript. We then repeat-masked these sequences 
(the masking parameters were left at the default values provided by Ensembl) and all the subsequences
corresponding to known coding exons. As a final result we obtained two lists of promoter regions, 
respectively including 21316 promoters for human and 21814 for mouse protein-coding genes.\\

\emph{miRNA genes.}~~ Following the idea discussed above that miRNAs belonging to the same TU are 
co-transcribed and thus should be co-regulated we chose to associate to all the pre-miRNAs belonging 
to a given TU the promoter  of the most 5'-upstream member of the TU (which is conventionally assumed as 
the TSS of the TU). This rule becomes trivial for single/isolated miRNAs. For each TU and isolated
miRNA we selected the promoter regions applying the following rules:

\begin{itemize}

\item if the pre-miRNA was non-genic, we selected the region ranging from nt -900 upstream to nt +100 downstrem
of the 5'-start of the pre-miRNA.
 
\item if the pre-miRNA was genic, with the same orientation of the host gene, we used the promoter region 
selected for the host gene.

\item if the pre-miRNA was genic, but with opposite orientation with respect to the host gene, 
we again selected the region ranging from nt -900 upstream to nt +100 downstream of the 5'-start of the pre-miRNA.

\end{itemize}

In all these cases, we then Repeat-Masked and exon-masked the sequences as we did for the protein-coding 
genes discussed above. Repeat-Masking was performed with the default values provided by Ensembl.

Merging together protein-coding and miRNA promoters we ended up with a collection of 21446 human and 21944
mouse regulatory sequences. 

\subsection*{Definition of 3'-UTR regions}

For the analysis of post-transcriptional regulation, we downloaded the complete 3'-UTR sequences 
for all protein-coding genes from the Ensembl database, version 46. Similarly to the promoters, 
we retained only those genes labeled as ``KNOWN''. Then we selected only the longest transcript, again 
among those labeled as ``KNOWN''. Since in the Ensembl database not all the genes have defined 3'-UTR 
regions we ended up with only 17486 human and 15921 mouse genes. We then Repeat-Masked these sequences 
using the default values provided by Ensembl as masking parameters.

It is worth noticing that, differently from the promoter case, the 3'-UTR sequences have different 
sizes. The average length of human or mouse 3'-UTR regions was respectively $\sim$1157 nts or $\sim$982 nts. 
    
\subsection*{Oligos analysis}

All the details relative to the oligos analysis are described in the Supporting Text, while the promoter and 3'-UTR 
sequences used are available in the Supporting website. The software implementation of the described algorithms is available 
upon request from the authors. 

\subsection*{TF-miR pairs and their joint target genes}

By crossing the lists of putative TF and miRNA targets obtained above
we constructed all possible feed-forward circuits composed by a transcription factor which
regulates a miRNA with which it coregulates a set of target genes.
In some cases in which a mature miRNA is transcribed from more 
than one genomic locus, all possible promoters were taken into account.

\subsection*{Assessment of miRNA targets using existing databases}

In silico predicted targets were obtained from the following three resources:
TargetScan, PicTar and miRBase. These three algorithms predict and assign target genes 
to miRNAs essentially based on sequence multi-species conservation. TargetScan targets 
were obtained from miRGen Release 3 (\verb+http://www.diana.pcbi.upenn.edu/miRGen/v3/+) where 
human miRNA family targets predicted by TargetScanS were downloaded from the TargetScan
Release 4.2 download site (\verb+http://www.targetscan.org/+) and miRNA family names were expanded 
to include all family members. We downloaded PicTar targets from the UCSC hg17 database
where they were presented as the picTarMiRNA4Way track. miRBase predicted targets were 
downloaded from \verb+http://microrna.sanger.ac.uk/targets/v4/+. Since different resources use 
different genomic annotation sets, we maintained Ensembl as main namespace and map both 
Gene Symbol IDs and RefSeq IDs to Ensembl Gene IDs.

\subsection*{Comparison with ECRbase}

From ECRbase (http://ecrbase.dcode.org/) we downloaded the complete dataset of 
transcription factor binding sites (TFBSs) predictions in the CoreECR regions 
(al least 355 nts long with 77\% indentity) from the \verb+tfbs_coreEcrs.hg18mm8.v94.txt+ file. 

We mapped the predicted TFBSs stored in those databases onto our promoter regions according 
to genomic coordinates, for protein-coding and miRNA genes. To avoid mismatches due to different
masking and / or misannotations we assigned the binding of ECRbase TF to our gene only if the 
complete sequence contained in the ECRbase was present in our promoter sequence.

\subsection*{Gene Ontology analysis}

We dowloaded the Gene Ontology (GO) annotation DAGs from the GO website
(\verb+http://www.geneontology.org+) and gene product annotations from the Ensembl database, 
version 46. We always considered a gene annotated to a GO term if it was directly annotated
to it or to any of its descendants in the GO graph. We implemented an exact Fisher's test
to assess whether a certain set of genes could be enriched in a certain GO category as done
in our previous experiences \citep{cora:2004,cora:2005}. The Fisher's test gave us the probability
$P$ of obtaining an equal or greater number of genes annotated to the term in a set made of the same 
number of genes, but randomly selected. To account for multiple testing, in this work, only 
P-values $< 10^{-4}$ were reported. 

\subsection*{Identification of cancer related genes}

OncomiRs were obtained from \citep{OncomiRs:Kerscher} and \citep{OncomiRs:Zhang}. 
We obtained the complete working list of mutated genes causally implicated in 
cancer from the Cancer Gene Census catalogue (\verb+http://www.sanger.ac.uk/genetics/CGP/Census/+).
The list was annotated with information concerning chromosomal location, tumour types in which 
mutations were found, classes of mutation that contributed to oncogenesis and other genetic properties.
We considered as cancer-related  a circuit if it included at least one oncomiR or one gene listed 
in the Cancer Gene Census catalogue. The full lists of these circuits, provided with detailed propertie
s on cancer-related genes, is available in the Supplementary files S9 and S10.

\vskip1.0cm {\bf Acknowledgements.} 
We thank Paolo Provero, Ferdinando Di Cunto, Francesca Orso, Paolo Macchi and Alessandro Quattrone
for useful suggestions and discussions. We also thank Marco Consentino-Lagomarsino for discussions about 
network motifs. This work was partially supported by the Fund for Investments of Basic Research 
(FIRB) from the Italian Ministry of the University and Scientific Research, No. RBNE03B8KK-006. 

\clearpage

\bibliography{molbiosystems_biblio}
\bibliographystyle{plos}

\clearpage

{\bf Supporting website}

\vskip1.0cm

All the Supplementary files and raw data are available at:\\
\verb+http://personalpages.to.infn.it/~cora/circuits/index.html+.

\clearpage

{\bf Table legends}

\vskip1.0cm

{\it Table 1:}  {\bf Most relevant mixed Feed-Forward Loops obtained with the Gene Ontology filter.}
Mixed Feed-Forward Loops (FFL) assembled with the pipeline outlined in Figure 2 and 
characterized by enriched Gene Ontology functional
annotations. For each circuit, we report the circuit id (FFL id: $TF|miRNA$) and the complete list of Joint Targets (JTs).
We then report some of the most relevant Gene Ontology annotations, with the relative P-values
evaluated by using a Fisher's test.
The complete dataset of circuits with their relative annotation is reported in Supplementary File S8.
Mature microRNA ids are written according to the standard nomeclature of miRBase \citep{Griffiths-Jones:2006}, 
for the TF and JT protein-coding genes, we used the standard HGNC ids.

\vskip1.0cm

{\it Table 2:}  {\bf Summary of mixed Feed-Forward Loops external annotations and relative examples.}
{\bf a)} General view: here we report the number of circuits presented in our database that
obtained the same number of external annotations, from 1 to 3. Detailed view: here we specify
the multiple external resources used for the annotation scheme and their relative contributions.
We report the number of circuits with assessed
link between: the transcription factor (TF) and the miRNA [TF$\rightarrow$miR]; the TF and a Joint Target
(JT) protein-coding gene [TF$\rightarrow$JT]; the mature microRNA (miR) and a JT [miR$\rightarrow$JT].
{\bf b)} Selection of a few circuits validated by the above tests.
The complete dataset of circuits is reported in Supplementary File S8.
For each circuit, we report the circuit id (FFL id: $TF|miRNA$) and the complete list of Joint Targets (JTs).
Mature microRNA ids are written 
according to the standard nomeclature of miRBase \citep{Griffiths-Jones:2006},
for the TF and JT protein-coding genes, we used the standard HGNC ids.

\vskip1.0cm

{\it Table 3:} {\bf Top ten transcription factors and microRNAs ranked by out-degree
and in-degree respectively.} Considering the links between transcription
factors (TF) and microRNA (miRNA) promoters defined in our transcriptional network,   
[TF$\rightarrow$miR link] we list the top ten TFs and miRNAs according to their   
out- and in-degree. The out-degree is defined, for a certain TF, as the
number of miRNAs directly controlled by the TF itself. The in-degree is
defined, for a certain miRNA, as the total number of TF acting on it.

\vskip1.0cm

{\it Table 4:} { \bf Cancer-related circuits.} Here we report the circuits which   
involve at least two cancer related items. For each circuit we indicated 
the circuit id (FFL id) in the first column, the master
transcription factor (TF) in the second column, the microRNA (miRNA) in the third column and 
the Joint protein-coding Target genes (JTs) in the fourth column. For each circuit only its cancer related items are listed
in the table, according to the role they serve within the circuit.
In the upper panel we report circuits in which the regulatory motifs
in the promoter regions of the miRNA and of the JTs
can be associated to a known TF. In the bottom panel
we report circuits for which the regulatory motif is uncharacterized. FFL id is
the identifier of a certain merged circuit, composed by the TF and miRNA names ($TF|miRNA$), or,
in case of unknown TF, by the exact DNA motif and the miRNA name.
Mature miRNA ids are
written according to the standard nomeclature of miRBase \citep{Griffiths-Jones:2006}, for the TF and JT protein-coding
genes, we used the standard HGNC ids.
For each circuit, the complete list of joint targets is available in Supplementary file S8.

\clearpage

{\bf Figure legends}

\vskip1.0cm

{\it Figure 1:} {\bf Feed-Forward Loops.} a) Representation of a tipical mixed
Feed-Forward Loop (FFL) analyzed in this work. In the square box, TF
is the master transcription factor; in the diamond-shaped box miR represents
the microRNA involved in the circuit while in the round box the Joint Target
is the joint protein-coding target gene (JT). Inside each circuit, $-\bullet$ indicates
transcriptional activation / repression, whilst $\dashv$ post-transcriptional
repression.
b) Flow-chart of the annotation strategies for the feed-forward circuits.
After building the catalogue of closed FFLs (see Figure 2), each side of the
circuit was expanded and analyzed using external support databases and
functional annotations. Beside each circuit link the source used
for its annotation is reported: see Materials and Methods for details.

\vskip1.0cm

{\it Figure 2:} { \bf Flow-chart of our pipeline for the identification of the
mixed Feed-Forward regulatory Loops.} We built two indipendent but
symmetrical pipelines for the construction of a transcriptional and, sepa-
rately, a post-transcriptional regulatory network in human. On the left: we
defined a catalogue of core promoter regions around the Transcription Start
Sites (TSS) for protein-coding and miRNA genes in the human genome. We
then applied a genome-wide sequence analysis strategy in order to identify,
a catalogue of human putative transcriptional regulatory motifs and the 
corresponding regulated genes. In so doing, the key
ingredients used were statistical properties of short DNA words (Oligo analysis) 
and conservation to mouse, implemented in an alignment-free manner
(conserved overrepresentation). On the right: a similar strategy was used,
starting from a catalogue of 3'-UTRs in human, to obtain a catalogue of human 
post-transcriptional regulated genes, with a focus for miRNA-mediated
interactions. We fixed 0.1 as FDR level for both the two motifs discovery pipelines.
At the end, the two regulatory networks were merged to extract
the complete dataset of closed mixed Feed-Forward Loops (FFLs), as defined in
Figure 1a and the results were filtered according to three different procedures:
by looking for (I) significant functional (Gene Ontology)
annotations between the Joint Targets of the FFLs, (II) independent computational evidences 
for the regulatory interactions of the FFLs, (III) relevance to cancer. 
See Materials and Methods for details.

\clearpage

{\it Figure 3:} {\bf Graphical representation of Type I and Type II circuits.} TF is the master transcription factor,
                miR represents the microRNA involved in the circuit and Joint Target is the joint target gene.
                Inside each circuit, $\rightarrow$ indicates transcription activation, whilst $\dashv$ transcription or
                post-transcriptional repression. In representing Type I and Type II circuits, we followed the
                nomenclature used in \citep{tsang}.

\vskip1.0cm

{\it Figure 4:} {\bf Graphical representation of the c-Myc$\mid$E2F1$\mid$hsa-miR-20a circuit, with its extension to E2F2}.
                The c-Myc$\mid$E2F1$\mid$hsa-miR-20a
                is the only feed-forward circuit already validated experimentally as stated by the literature.
                Its components are embedded in a more sophisticated network, in particular mining our database we
                recognized the interplay with E2F2. E2F2 is downregulated by hsa-miR-20a at the post-transcriptional level, and it is a direct
                transcriptional target of E2F1 itself. $-\bullet$ indicates 
		transcriptional activation / repression, whilst $\dashv$ post-transcriptional repression. Mature microRNA ids are
		written according to the standard nomeclature of miRBase \citep{Griffiths-Jones:2006}, for the TF and JT protein-coding
		genes, we used the standard HGNC ids.
\vskip1.0cm

{\it Supplementary Figure S1:} {\bf Randomization results for the network motifs analysis of mixed feed-forward loops.}        
We plotted the number of single-target mixed Feed-Forward Loops (FFLs)
obtained in the real network and associated to known Transcription Factors (blue line)
alongside the distributions (normalized histograms) of the number of single-target
mixed FFLs detected for the three randomization strategy adopted. The red data refer
to the results obtained with the {\it Random reshuffling of microRNA promoters and seeds}, 
whilst the light green refers to the {\it Edge switching } randomization strategy and the dark green to the 
{\it Complete node replacement} one.
The figure is divided into two separated panels: panel A) contains
results relative to the (-900/+100) nts window for the definition of promoter of miRNAs
and protein-coding genes, whereas panel B) contains results relative to the (-500/+100)
nts case. See Supporting Text for details.

\newpage

{\bf List of Supplementary files}

\vskip1.0cm

{\it Supplementary file 1:} {\bf List of human and mouse pre-miRNAs included as transcriptional units (TU) representatives.}
\vskip0.5cm
{\it Supplementary file 2:} {\bf List of human and mouse mature miRNAs included in our study.}
\vskip0.5cm
{\it Supplementary file 3:} {\bf Transcriptional regulatory network: conserved-overrepresented oligos in promoter regions of protein-coding and microRNA 
			    genes and their association to known transcription factor binding sites.}
\vskip0.5cm
{\it Supplementary file 4:} {\bf Transcriptional regulatory network: conserved-overrepresented oligos in promoter regions of microRNA genes and 
			    their association to known transcription factor binding sites.}
\vskip0.5cm
{\it Supplementary file 5:} {\bf Post-transcriptional regulatory network: conserved-overrepresented oligos in 3'-UTRs of protein-coding genes and 
			    their association to known microRNA seeds.}
\vskip0.5cm
{\it Supplementary file 6:} {\bf Raw data of circuits.}
\vskip0.5cm
{\it Supplementary file 7:} {\bf Raw data of circuits related to unknown Transcription Factors.}
\vskip0.5cm
{\it Supplementary file 8:} {\bf Circuits final datasets with complete annotations.}
\vskip0.5cm
{\it Supplementary file 9:} {\bf Circuits relevant to cancer, related to known Transcription Factors.}
\vskip0.5cm
{\it Supplementary file 10:} {\bf Circuits relevant to cancer, but related to unknown Transcription Factors.}
\vskip0.5cm
{\it Supplementary file 11:} {\bf Raw data of circuits for the alternative choice of the promoter region: (-500/+100) around the TSS.}

\clearpage

\begin{center}
\vskip 0.7cm
{\Large  Supporting Information}
\end{center}

\setcounter{footnote}{0}
\def\thefootnote{\arabic{footnote}}

\vskip2.0cm
\section*{Oligos analysis}

We searched for statistically significant motifs in  promoters and 3'-UTRs using the algorithms developed
in our previous works \citep{cora:2005,cora:2007}, with minor modifications. Here we report the main 
features of these algorithms for completeness.\\

\emph{Motif search in the promoter regions}

We preprocessed the promoter regions by merging
overlapping fragments, in order to build a non-redundant
set of sequences for the evaluation of the background probabilities.

Both the original dataset of promoter regions and the set of non overlapping
blocks were then separated according to their CG content. Following what we did in
ref~\citep{cora:2007}
we associated to each gene a label (CG-rich or CG-poor) based on the CG content of the
corresponding promoter region using the median of the CG content of the whole set of promoter regions as
threshold. \\

\emph{Motif search in the 3'-UTR regions}

We applied the same pipeline to the set of 3'-UTRs. We first built a set of
non-redundant sequences by merging the
overlapping ones and then divided both datasets (the original and the non-redundant one) into CG-rich and CG-poor
subsets. The only difference with respect to the promoter case was that we merged the sequences only when they
were located on the same strand. \\

Once equipped with those dataset, we constructed, separately for human and mouse, the sets $S(w)$ of genes such
that the oligo $w$ is overrepresented in the corresponding promoter or 3'-UTR. This was done for all 6 to 9 nt oligos
for promoter regions and for 7 nt oligos for 3'-UTRs through the following steps:
    \begin{itemize}

    \item We computed the overall frequency $f(w)$ as the ratio
      \begin{equation}
        f(w) = \frac{N(w)}{N}
      \end{equation}
      where $N(w)$ is the number of times $w$ occurs in the collection of all
     the non-redundant sequences, and $N=\sum_w N(w)$. 

    \item For each gene $g$ let $n_g(w)$ be the number of occurrences of $w$
      in the original (i.e. before merging overlapping sequences)
       promoter or 3'-UTR of $g$. We computed the overrepresentation Pvalue as
      \begin{equation}
        P_g(w) = \sum_{k=n_g(w)}^{n_g}\left(n_g\atop k \right) f(w)^{k}
        (1-f(w))^{n_g-k}
      \end{equation}
      where
      \begin{equation}
        n_g = \sum_w n_g(w)
      \end{equation}
      is the total number of oligos of the same length as $w$ that can be read
      in  the promoter or the 3'-UTR of $g$. Self-overlapping matches of the same oligo
      were discarded \citep{vanhelden:1998}. Motifs were counted on both strands in the promoter case and 
      only on the transcribed strand in the 3'-UTR case.

    \item The genes for which $P_g(w) < 0.01$ were included in the set $S(w)$. Notice that no biological significance was ascribed 
          to these sets before they are selected for evidence of evolutionary conservation as explained later: 
          therefore the choice of the cutoff on P can be arbitrarily lenient and in particular no correction for multiple testing
          was applied in this step. 

    \end{itemize}

    The procedure described above was performed separately for CG-rich and
    CG-poor genes, so as to identify overrepresented words with respect to the
    appropriate background frequencies. For each word $w$ the sets $S(w)$ computed for CG-rich
    and CG-poor genes were then joined to obtain a single set $S(w)$. This procedure was performed separately for
    human and mouse. \vskip2.0cm

\subsection*{Conservation of overrepresentation}  

    We define an oligo $w$ as
    ``conserved overrepresented'' if the sets of genes $S_{human}(w)$ and
    $S_{mouse}(w)$ contain a significantly larger number of orthologous
    genes than expected by chance. Pairs of human-mouse orthologous genes
    were obtained from Ensembl, selecting only orthologous defined as Unique
    Blast Reciprocal Hit to obtain one-to-one orthology
    relationships. For miRNA genes, orthology relatioships were downloaded 
    from \citep{landgraf}.

    Let $M$ be the total number of human genes represented in our sequences
    which have a mouse ortholog. Given an oligo $w$ and the set
    $S_{human}(w)$, let $m$ be the number of human genes in $S_{human}(w)$
    which have a mouse ortholog, $N$ the number of genes in $S_{mouse}(w)$
    with a human ortholog, and $n$ the number of genes in
    $S_{human}(w)$ with a mouse ortholog in $S_{mouse}(w)$. We then
    compute the Pvalue
    \begin{equation}
      P = \sum_{k=n}^{m} F(M,m,N,k)
    \end{equation}
    where
    \begin{equation}
      F(M,m,N,k) = \frac{\left(m \atop k\right) \left(M - m \atop N -
          k\right)}{\left(M \atop N\right)}
    \end{equation}

    Multiple testing was taken into account with a Benjamini-Yekutiely FDR approach~\citep{by2001},
    and conserved overrepresentation was defined to be significant when the
    Benjamini-Yekutiely corrected Pvalues were less than 0.1. Note that a similar approach
    was used e.g. in \citep{chan:2005}, and termed ``Network conservation''; basically the main difference with respect to our 
    approach is that in \citep{chan:2005} the simple presence of a k-mer is used instead of statistical overrepresentation. 

\subsubsection*{Search of putative TF binding sequences in the promoter motifs database}

We searched for putative TF binding sequences in the set of conserved overrepresented
promoter motifs with two different and complementary approaches.

\begin{itemize}
\item {\bf TRANSFAC analysis}

We performed a weight matrix search of TF binding sites among our motifs using the Match
web-based tool that is integrated in TRANSFAC Professional (release 11.2). In a few cases the motif
turned out to be compatible with more than one weight matrix. In these cases we associated all
the potential binding factors to the motif. Weight matrix profiles adopted in our search were limited
to the vertebrate-specific subset of high quality matrices with predefined cutoffs for core and
matrix similarity optimized in order to minimize the sum of both false positive and false negative rates.

To restrict the search, we specified the predefined human site set and allowed no mismatching
nucleotide in a match between search pattern and motif.

\item {\bf Comparison with the Xie et al.\citep{xie:2005} list of binding sequences}

A second approach was the direct comparison
with the consensus sequences for vertebrate TFs published in \citep{xie:2005} (see Supplementary
table S3 of their work). The association between motif and TF was accepted only if our motif
exactly overlapped (according to the IUPAC alphabet) the consensus taken from the \citep{xie:2005} list.

\end{itemize}

In this way we could obtain a list of putative target genes for the TF's contained in TRANSFAC and in the
\citep{xie:2005} paper: for each TF we simply merged together the genes contained in the sets $S(w)$
associated to all the words $w$ compatible with the TRANSFAC weight matrix  or the \citep{xie:2005}
consensus.

\subsection*{Search of putative miRNA binding sequences in the 3'-UTR motifs database}

We obtained human mature miRNA sequences from the miRBase Sequence Database (release 9.2)
in flat file form from \verb+ftp://ftp.sanger.ac.uk/pub/mirbase/sequences/+. We followed standard
seed parameter settings and considered 7 nts long seeds, beginning within the third position from
the miRNA 5' end (nt 1 or 2 or 3). We identified a motif (a miRNA binding site) if
it had a perfect Watson-Crick match with at least one of the possible miRNA seeds~\citep{cora:2007}.

In this way we could obtain a list of putative target genes for each one of the 193 mature miRNAs
contained in our database: for each miRNA we merged together the genes contained in the sets $S(w)$
associated to all the words $w$ identified as miRNA binding sites.

\section*{Randomizations for network motifs analysis}

We adopted three different randomization protocols to study the statistical properties of our  
 mixed FFLs. \\

{\bf Random reshuffling of miRNA promoters and seeds} 
We constructed a randomized versions of miRNA promoters and seeds and then constructed the putative FFLs following 
the same pipeline discussed in the text.
All the other parameters of the pipeline were left unchanged and we performed a total of
50 randomizations and evaluated the mean value and standard deviation of the number of FFLs. 
For each randomization, the catalogue of 
random miRNA promoters was build performing $ N log N $ swaps between two randomly selected nucleotides in the original
promoter sequence, preserving the repeat-masking (being $ N $ the effective length of the miRNA promoter sequence). 
In this way, the CG content of each miRNA promoter was kept unchanged. To build the set of random 7-mers used as miRNA seeds,
we started with the catalogue of real seed regions used in this work and performed, as above, $ N log N $ swaps between two randomly 
selected nucleotides, and retaining only oligos not already identified with any known miRNA seed. \\

{\bf Edge switching}, 
(see \citep{martinez:2008}). In this type of randomization, we started from the  transcriptional and post-transcriptional regulatory 
networks obtained with our sequence analysis pipeline (and reported in Supplementary files S3 and S5), and applied the following 
randomization strategy: we randomly selected two TFs (or miRNAs) and two of their target genes, one for each of them. We then
swapped the two target genes, only if the selected genes were not already present in the list of regulated genes of the two TFs 
(or miRNAs). In that way, the out-degree of a certain TF and the in-degree of a certain target gene were preserved, whilst 
the connections were randomized; the same for a miRNA and its targets. For each complete randomization cycle, a total of $10^{6}$   
exchanges was executed. The randomization protocol was 
performed 1000 times and the mean value and standard deviation of the number of FFLs was evaluated.\\

{\bf Complete node replacement}, (see again \citep{martinez:2008}). In this type of randomization, we started again from our
 transcriptional and post-transcriptional 
regulatory networks, and applied a less constrained randomization strategy: we randomly selected 
two TFs (or miRNAs) and two of their target genes, one for each of them. We then swapped the two target genes, without any filter.
In this way, the global network topology of the original regulatory network was completely modified. For each complete 
randomization cycle, a total of $10^{6}$ exchanges was executed. Also this randomization protocol was 
performed 1000 times and the mean value and standard deviation of the number of FFLs was then evaluated.\\

\newpage

{\it Supplementary Figure S1:} {\bf Randomization results for the network motifs analysis of mixed feed-forward loops.}
We plotted the number of single-target mixed Feed-Forward Loops (FFLs)
obtained in the real network and associated to known Transcription Factors (blue line)
alongside the distributions (normalized histograms) of the number of single-target
mixed FFLs detected for the three randomization strategies adopted. The red data refer
to the results obtained with the {\it Random reshuffling of microRNA promoters and seeds},
whilst the light green refers to the {\it Edge switching } randomization strategy and the dark green to the
{\it Complete node replacement} one.
The figure is divided into two separated panels: panel A) contains
results relative to the (-900/+100) nts window for the definition of promoter of miRNAs
and protein-coding genes, whereas panel B) contains results relative to the (-500/+100)
nts case. See Supporting Text for details.

\clearpage
\thispagestyle{empty}

\begin{figure}[!ht]
  \centering
  \includegraphics[scale = 0.70 , angle = 90 ]{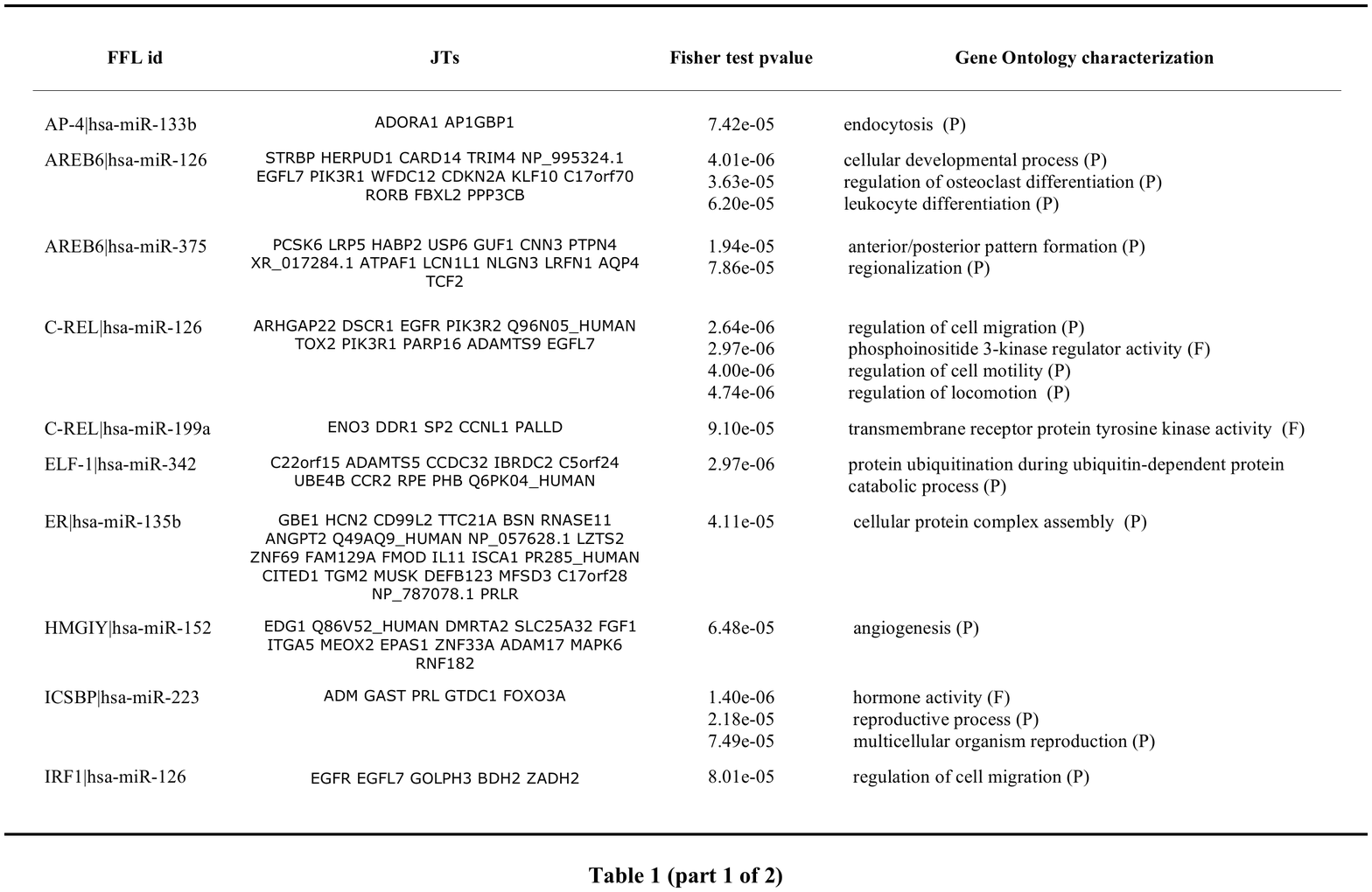}
\end{figure}

\clearpage
\thispagestyle{empty}

\begin{figure}[!ht]
  \centering
  \includegraphics[scale = 0.70 , angle = 90 ]{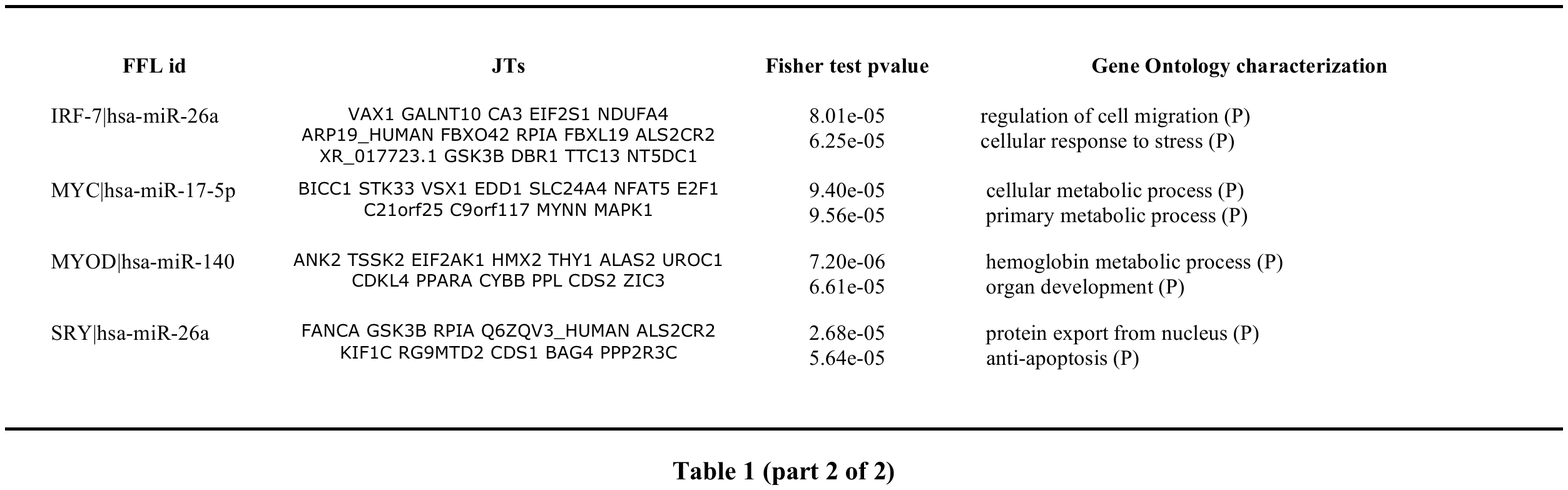}
\end{figure}

\clearpage
\thispagestyle{empty}

\begin{figure}[!ht]
  \centering
  \includegraphics[scale = 0.70 ]{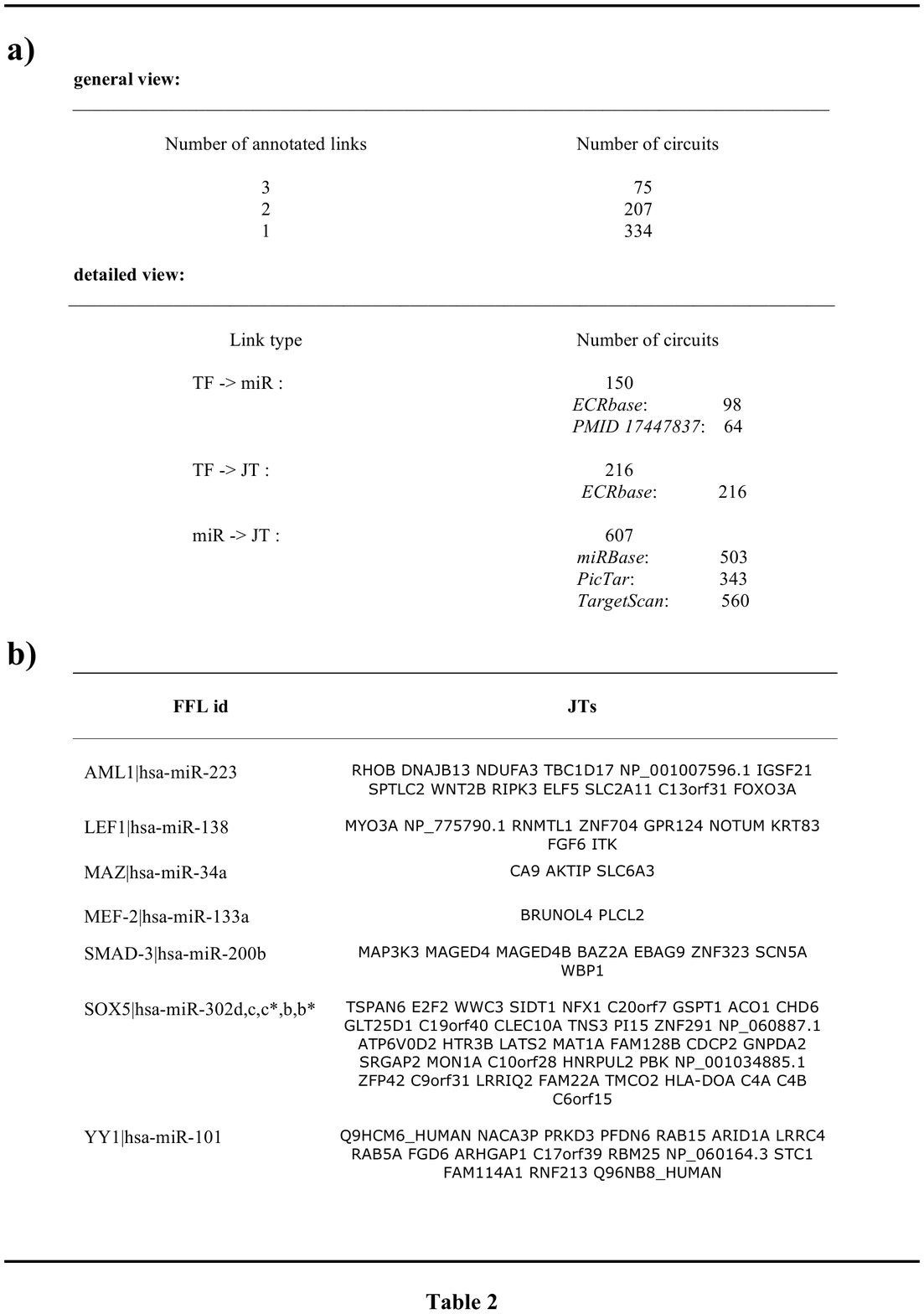}
\end{figure}

\clearpage
\thispagestyle{empty}

\begin{figure}[!ht]
  \centering
  \includegraphics[scale = 0.70 ]{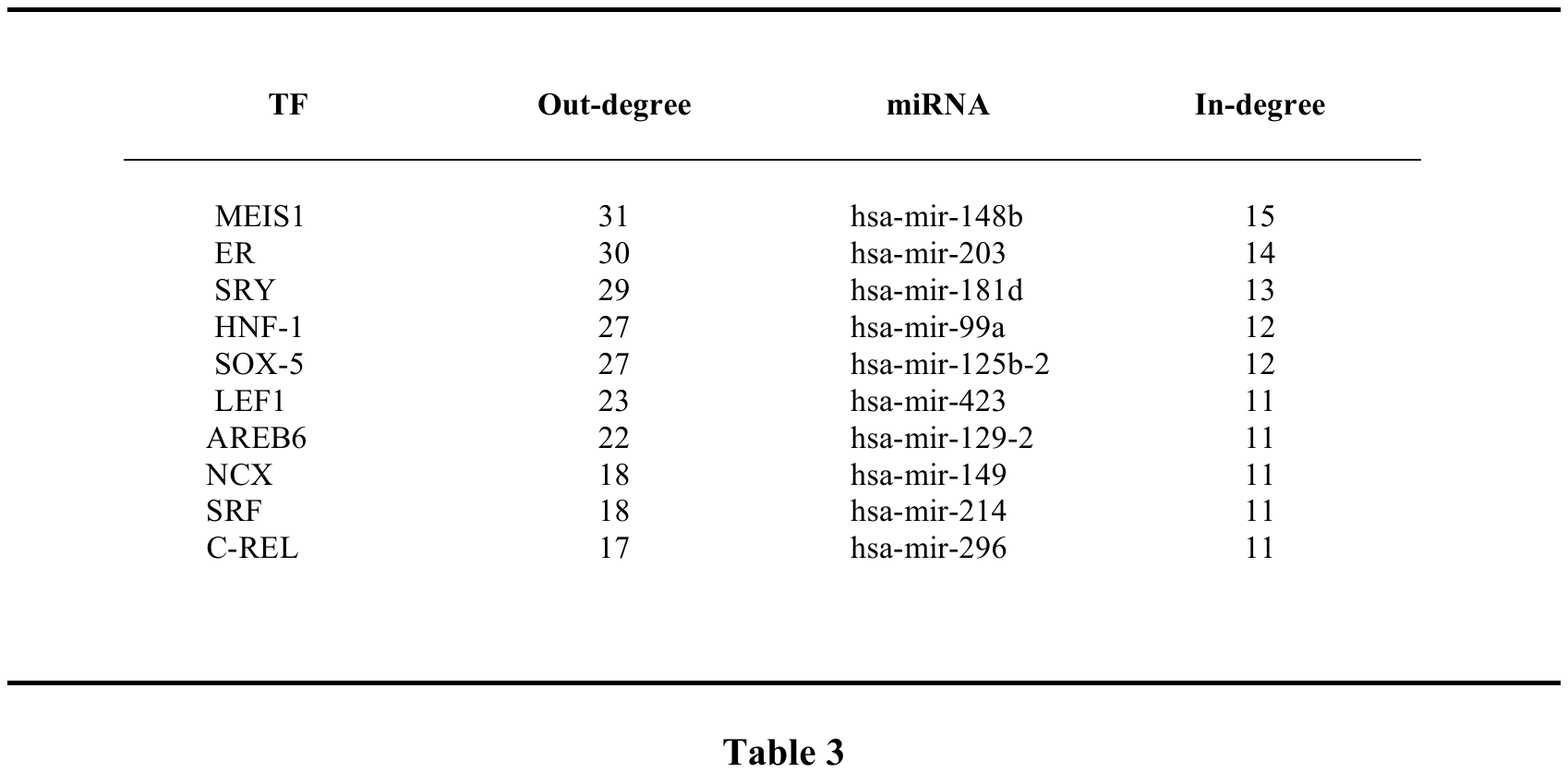}
\end{figure}

\clearpage
\thispagestyle{empty}

\begin{figure}[!ht]
  \centering
  \includegraphics[scale = 0.70 ]{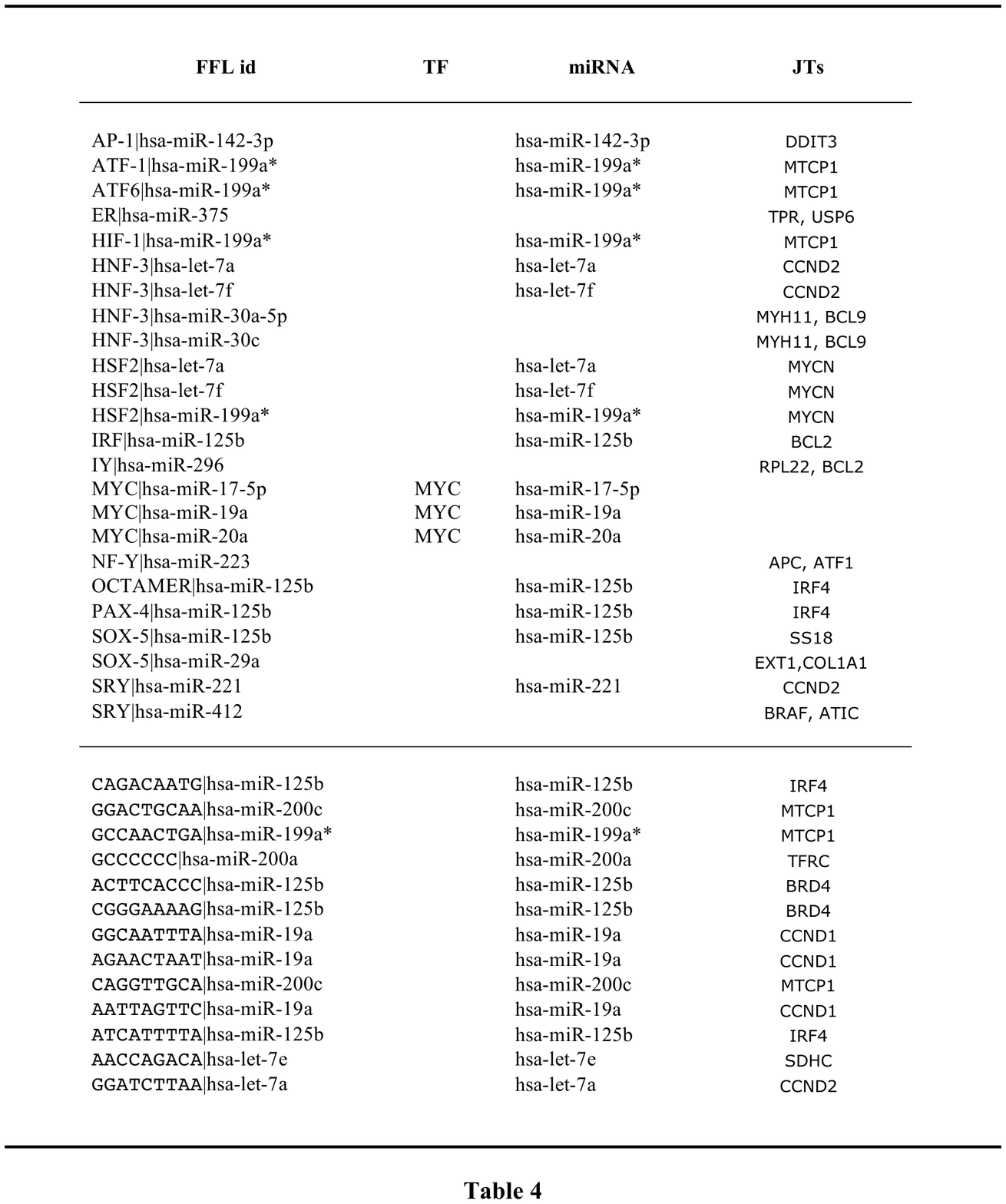}
\end{figure}

\clearpage
\thispagestyle{empty}

\begin{figure}[!ht]
  \centering
  \includegraphics[scale = 0.60 ]{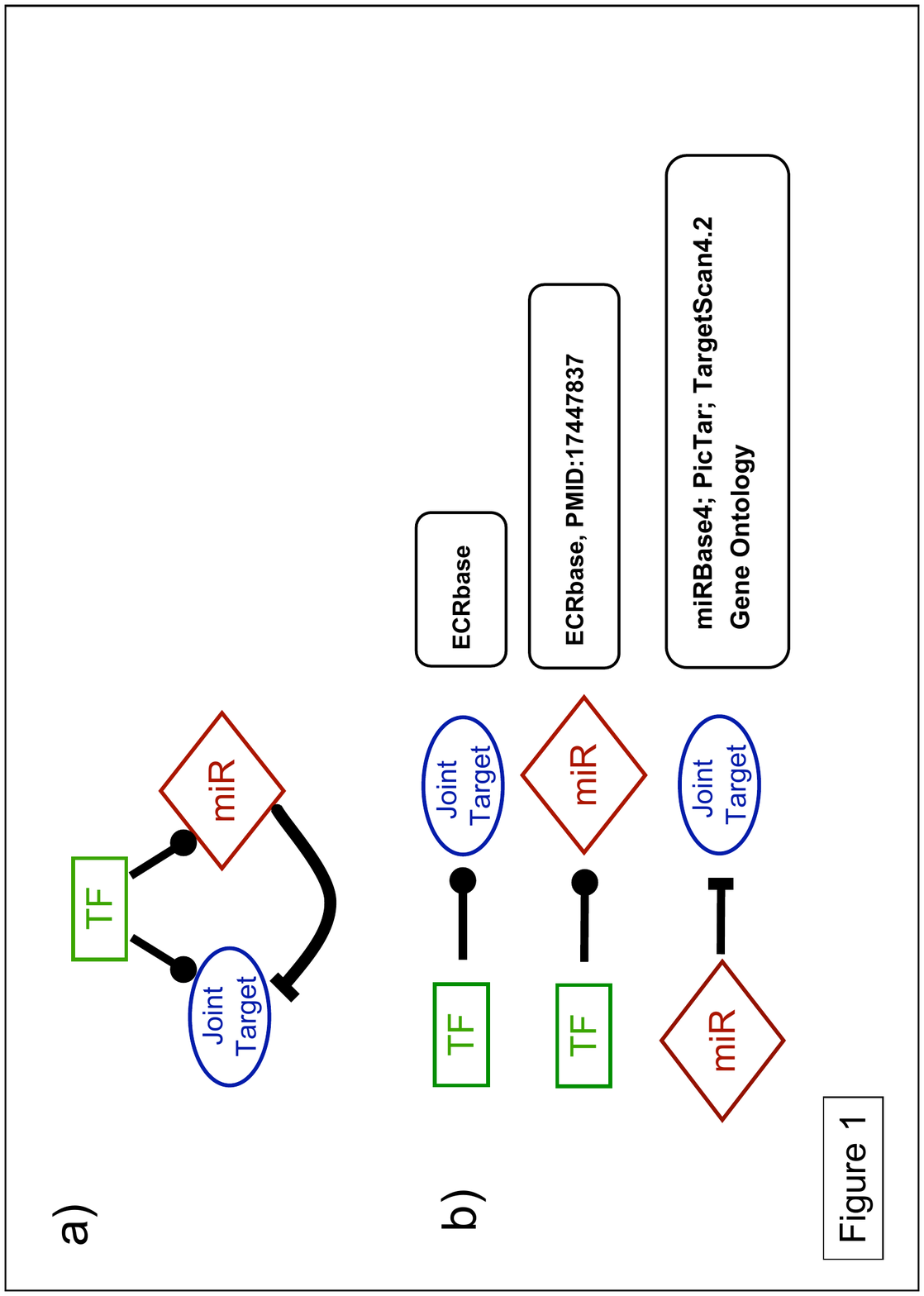}
\end{figure}

\clearpage
\thispagestyle{empty}

\begin{figure}[!ht]
  \centering
  \includegraphics[scale = 0.60 ]{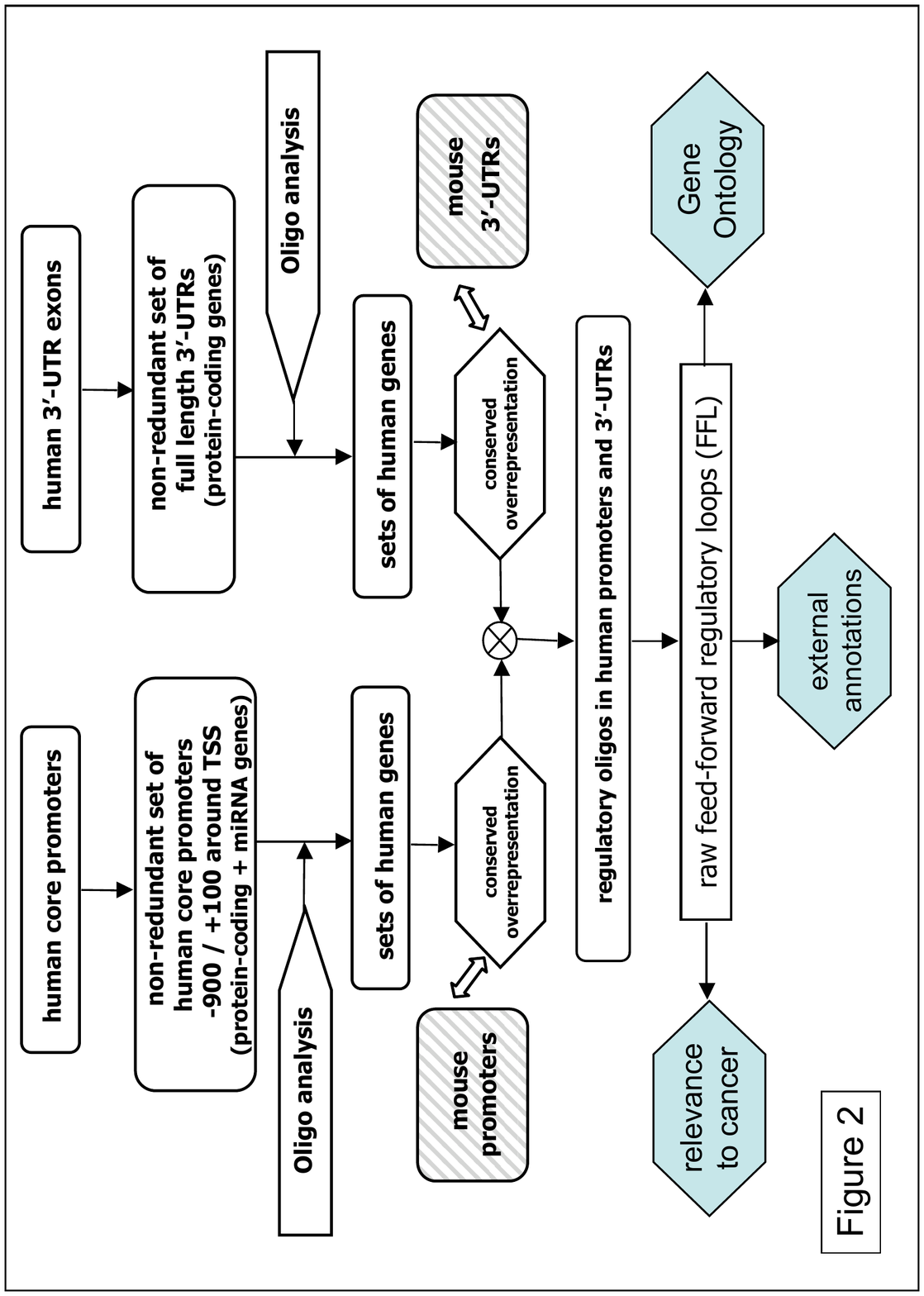}
\end{figure}

\clearpage
\thispagestyle{empty}

\begin{figure}[!ht]
  \centering
  \includegraphics[scale = 0.60 ]{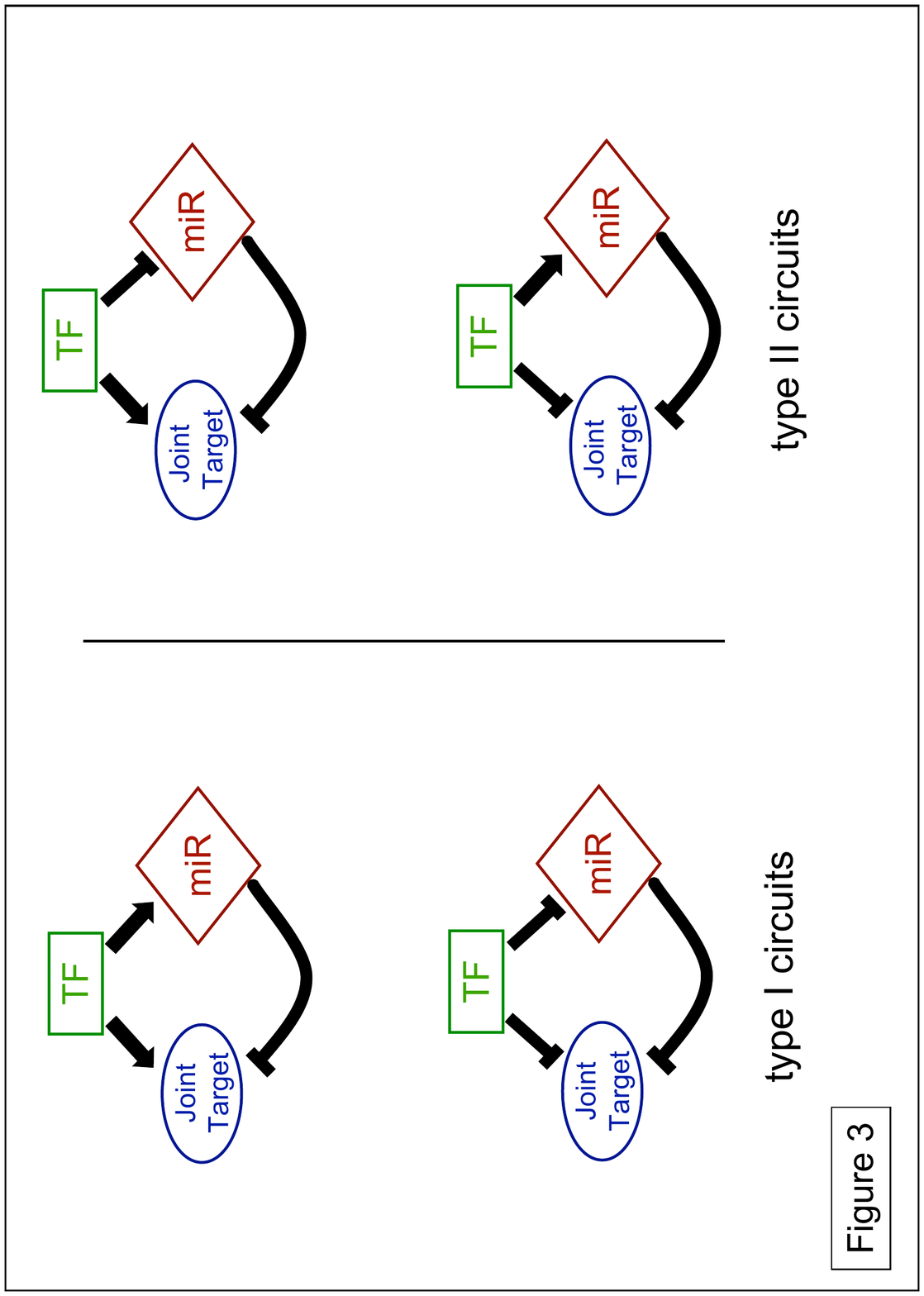}
\end{figure}

\clearpage
\thispagestyle{empty}

\begin{figure}[!ht]
  \centering
  \includegraphics[scale = 0.70 ]{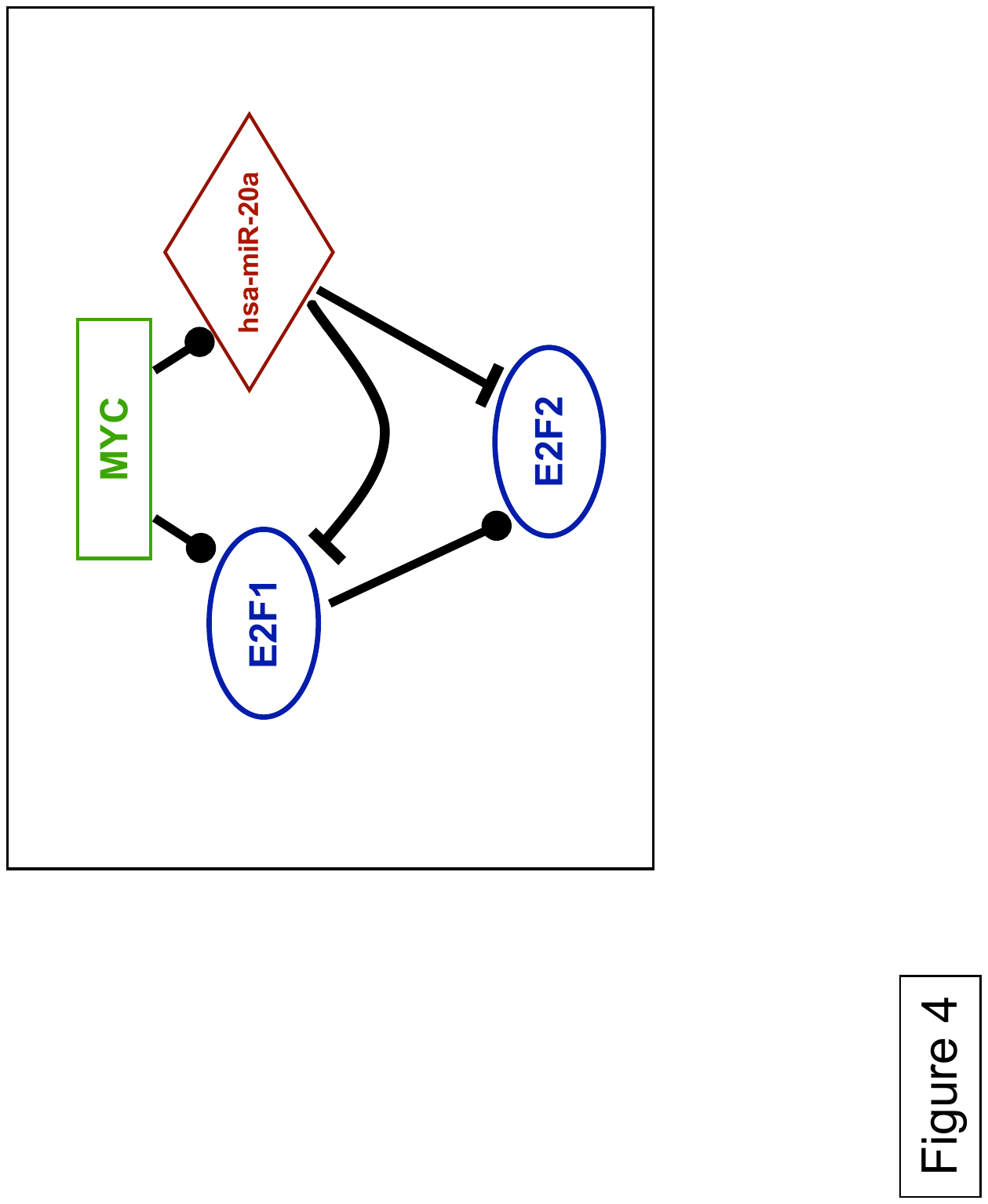}
\end{figure}

\clearpage  
\thispagestyle{empty}

\begin{figure}[!ht]
  \centering
  \includegraphics[scale = 0.70 ]{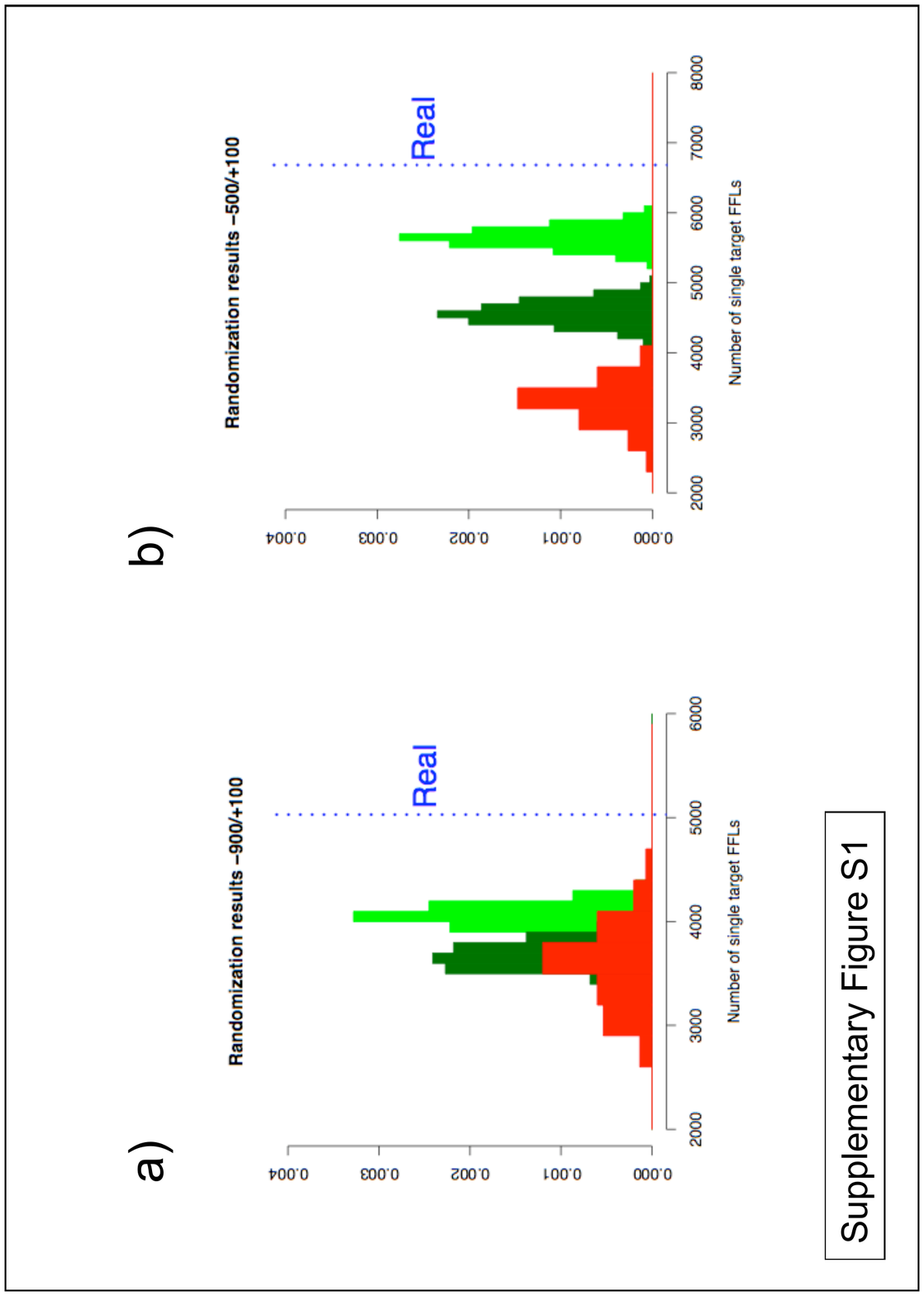}
\end{figure}

\end{document}